\documentclass[aps,pra,reprint,superscriptaddress,amsmath,showpacs,amsfonts,amssymb]{revtex4-1}

\usepackage{graphicx}
\usepackage{bm}
\usepackage{color}
\usepackage{MnSymbol}
\usepackage{enumerate}
\usepackage{bbold}
\usepackage{lipsum}
\usepackage{array}
\DeclareMathAlphabet{\mathpzc}{OT1}{pzc}{m}{it}
\usepackage{tikz}
\usetikzlibrary{shapes,arrows}
\usetikzlibrary{positioning,arrows}
\usetikzlibrary{decorations.pathmorphing}
\usetikzlibrary{decorations.markings}

\tikzset{z/.default=0} 
\tikzset{x/.default=0} 

\tikzstyle{line} = [draw, -latex']
\tikzstyle{line2} = [draw,thick]
\tikzstyle{s} = [circle,draw,fill=yellow!20,node distance=1cm,scale=0.7]
\tikzstyle{z} = [circle,draw,fill=red!20,node distance=1cm,scale=0.7]
\tikzstyle{x} = [circle,draw,fill=blue!20,node distance=1cm,scale=0.7]
\tikzstyle{t} = [circle,draw,node distance=1cm,scale=0.7]
\tikzstyle{o} = [draw,node distance=1cm,scale=0.9]
\tikzstyle{ov} = [draw,node distance=1cm,scale=0.9]
\tikzstyle{b} = [node distance=1cm,scale=1]
\tikzstyle{test} = [style={path picture={\draw[black](path picture bounding box.south east) -- (path picture bounding box.north west) (path picture bounding box.south west) -- (path picture bounding box.north east);}},node distance=0cm,scale=0.5]

\def\la{\langle}
\def\ra{\rangle}

\def\be{\begin{equation}}
\def\ee{\end{equation}}

\newcommand{\op}[1]{\hat{ #1}}                

\newcommand{\td}[1]{\tilde{ #1}}

\newcommand{\Tr}[1]{\text{Tr}(#1)}               

\newcommand{\dd}{\mathrm{d}}

\newcommand{\nn}{\nonumber}
\newcommand{\pp}{{\bm p}}
\newcommand{\qq}{{\bm q}}
\newcommand{\bxx}{X}
\newcommand{\bxt}{\tilde{X}}
\newcommand{\bjj}{J}
\newcommand{\bjt}{\tilde{J}}
\newcommand{\xp}{u}
\newcommand{\yp}{v}
\newcommand{\zp}{w}
\newcommand{\pxp}{p_u}
\newcommand{\pyp}{p_v}
\newcommand{\pzp}{p_w}
\newcommand{\diel}{\lambda}
\newcommand{\sech}{\text{sech}}

\begin{document}
\tikzset{
particle/.style={thick,draw=black, postaction={decorate},decoration={markings,mark=at position .5 with {\arrow[draw]{<}}}},
particle2/.style={thick,draw=black, postaction={decorate},decoration={markings,mark=at position .5 with {\arrow[draw]{<}}}},
gluon/.style={thick,decorate, draw=black,decoration={coil,aspect=0,segment length=3pt,amplitude=1pt}}
 }

\title{Stochastic path integral formalism for continuous quantum measurement}

\author{Areeya Chantasri}
\affiliation{Department of Physics and Astronomy, University of Rochester, Rochester, NY 14627, USA}
\affiliation{Center for Coherence and Quantum Optics, University of Rochester, Rochester, NY 14627, USA}
\author{Andrew N. Jordan}
\affiliation{Department of Physics and Astronomy, University of Rochester, Rochester, NY 14627, USA}
\affiliation{Center for Coherence and Quantum Optics, University of Rochester, Rochester, NY 14627, USA}
\affiliation{Institute for Quantum Studies, Chapman University, 1 University Drive,
Orange, CA 92866, USA}

\date{\today}

\begin{abstract}
We generalize and extend the stochastic path integral formalism and action principle for continuous quantum measurement introduced in [A. Chantasri, J. Dressel and A. N. Jordan, Phys. Rev. A {\bf 88}, 042110 (2013)], where the optimal dynamics, such as the most likely paths, is obtained by extremizing the action of the stochastic path integral. In this work, we apply exact functional methods as well as develop a perturbative approach to investigate the statistical behaviour of continuous quantum measurement. Examples are given for the qubit case. For qubit measurement with zero qubit Hamiltonian, we find analytic solutions for average trajectories and their variances while conditioning on fixed initial and final states. For qubit measurement with unitary evolution, we use the perturbation method to compute expectation values, variances, and multi-time correlation functions of qubit trajectories in the short-time regime. Moreover, we consider continuous qubit measurement with feedback control, using the action principle to investigate the global dynamics of its most likely paths, and finding that in an ideal case, qubit state stabilization at any desired pure state is possible with linear feedback. We also illustrate the power of the functional method by computing correlation functions for the qubit trajectories with a feedback loop to stabilize the qubit Rabi frequency.
\end{abstract}
\maketitle
\date{today}

\section{Introduction}
Continuous measurement of quantum systems \cite{BookNielsen,BookBreuer}, the study of quantum systems states under the influence of observation prolonged in time, has been a topic of considerable activity in recent years. Particularly, for the measurement of an individual microscopic system that is weakly coupled to measurement apparatus, the system's state and its conditioned evolution in time, the so-called diffusive-type quantum state trajectory \cite{BookCarmichael,BookWiseman,Wiseman1996,Jacobs2006}, have been intensively explored for applications in quantum information and quantum control. Some of the active fields are quantum state estimation (i.e., estimating the pre-measurement state of an ensemble of identically prepared systems \cite{Verstraete2001,*Smith2006,*Gammelmark2014}), conditional reversal of measurement \cite{Korotkov2006,*Korotkovexp2008,*Lange2014}, and the preparation of entangled states \cite{Rusko2003,*Roch2014,*Riste2013}. The ability to continuously measure quantum systems also opens the possibility of feedback control \cite{Wiseman1994,*Doherty2000,*Ibarcq2013}, which has also been investigated for topics such as the stabilization of coherent oscillations \cite{Ruskov2002,Korotkov2005,Korotkov2005-2,Korotkov2012} and rapid state purification \cite{Jacobs2003,*Combes2006,*Combes2008,*Jordan2006,*Wiseman2006,*Wiseman2008,*Ruskov2012}.


This growing interest in the quantum systems under weak continuous measurement has motivated a thorough analysis of quantum trajectory statistics. Of notable importance are advances in experiments, such as the measurement of superconducting qubits \cite{Gambetta2008,Kater2013}, which has allowed the tracking of the trajectories of the quantum state with high fidelity in a single measurement run, allowing the statistics of selected subensembles of trajectories to be explored. The authors recently developed an action principle \cite{Chantasri2013} over a doubled quantum state space, based on a path integral representation of probability distributions of quantum trajectories. The action principle, implemented by extremizing of the stochastic path integral's action, was used to investigate the optimal behaviour of the trajectories with arbitrary constraints, such as fixing the final boundary condition \cite{Weber2014,Silveri2015}. The stochastic path integral and the optimum likelihood approach provide a convenient way to investigate statistical distributions and globally optimal dynamics of quantum state evolution, in addition to the stochastic master equations describing the quantum trajectories and the Lindblad master equations describing their average evolution \cite{Lind1976,BookCarmichael}.

In this paper, we continue the development of the stochastic path integral formalism \cite{Chantasri2013}, to further explore advantages of having the full joint probability distribution of quantum trajectories. This includes computing statistical averages or expectation values with the ability to condition on definite quantum states at particular times. We present several examples of the formalism including a qubit system under the influence of measurement alone, measurement with concurrent unitary dynamics, and qubit measurement with feedback control. In these examples, the statistics of qubit trajectories are computed using developed techniques for the path integral, such as multi-dimensional Gaussian integrals and diagrammatic expansion theory. Moreover, in an example of qubit measurement with linear feedback, we utilize a phase portrait analysis to investigate the most likely behaviour of the system dynamics, revealing a simple and practical approach for qubit state stabilization.

There have been past works on continuous quantum measurements with path integrals, so we wish to discuss how our approach bears both similarities and differences to them. An early approach suggested by Feynman \cite{Feyn1948} and later independently developed by Mensky \cite{Mensky1979,*Mensky1994} is a restricted path integral: a modified version of the Feynman path integral to only sum over paths that contribute to a measurement record. Caves, and also Barchielli \cite{Caves1986,*Caves1987,CavesMilb1987,Barchielli1982}, constructed similar path integrals by adding coarse-graining (resolution) functions describing the effect of the measurement. In these path integral approaches, one considers the distribution of the measurement records (it can be derived from the probability amplitude, see Appendix~\ref{app-otherpathint}), whereas in our approach, we formulate a path integral in quantum state space to represent a joint probability distribution of the measurement readouts as well as quantum state trajectories. Wei and Nazarov discussed a different approach to continuous measurements using the Keldysh path integral technique \cite{Nazarov2008}. In Breuer and Petruccione's path integral \cite{Breuer1997,BookBreuer}, the notion of the sum over pure state paths in Hilbert space is applied, resulting in a different type of doubled state space that does not yield an action functional.

The stochastic path integral formalism and the analysis of its action are also applied in classical stochastic processes. For instance, the formalism is used in studying the dynamics and distribution of transmitted electronic charge \cite{Jordan2003,*Jordan2004,*JordanSuk2004,*SukhorJordan2007}, the neural network  \cite{Buice2007,*Buice2015}, and the large deviations from typical behaviours (rare events) \cite{Dykman1994,*Elgart2004,*Bernard2006,*Sinitsyn2009}. Our approach is similar to the Martin-Siggia-Rose formalism \cite{Martin1973}, which involves adding conjugate fields through the Fourier integral form of delta functionals enforcing diffusive dynamics. Notably, one can think of the quantum trajectories on a finite dimensional state space as analogous to classical random processes in configuration space of that dimension.

This paper is organized as follows. In section~\ref{sec-mainSPI}, we review the stochastic path integral formalism and its extremized action equations, for a general finite-dimensional system with a Markovian setup for weak continuous measurement. In section~\ref{sec-qubit}, a specific measurement setup for a qubit is presented, which is used throughout this paper. In section~\ref{sec-qnd}, we show that, in the case without qubit Hamiltonian, we can perform the full path integration directly to get the multi-time correlation functions for the preselected and postselected qubit state. In section~\ref{sec-purbexpand}, the diagrammatic expansion theory is presented as an alternative approximation method for computing multi-time correlation functions, with examples for qubit measurement with Rabi oscillation. In section~\ref{sec-feedback}, qubit measurement with feedback control and its optimal dynamics are investigated using the path integral and the action principle approaches. The conclusion can be found in section~\ref{sec-conclusion}. A series of supplementary discussions and some detailed calculations that are not included in the main text are presented in the Appendixes.

\section{Stochastic path integral and its optimal paths}\label{sec-mainSPI}
We consider the distribution of quantum state trajectories for continuous quantum measurement. A quantum state trajectory, or simply a quantum trajectory, is an evolution of a quantum state in time, conditioned on a detector readout realization. This conditional state trajectory is also known as a solution of stochastic master equations, unravelling master equations in Lindblad form. Let us discretize the measurement readout into $n$ time points and denote $\{ r_k \}_{k=0}^{n-1}$ as a measurement realization. Each $r_k$ is a readout obtained between time $t_k$ and $t_{k+1}=t_k +\delta t$, and is assumed dependent only on a quantum state right before its measurement (Markov assumption). We define a series of quantum states  $\{ \qq_k \}_{k=0}^{n}$, written as a $d$-dimensional parametrized vector $\qq$, where the components are the expansion coefficients of the density operator $\rho$ written in some orthogonal operator basis, such as the $N^2-1$ generalized Gell-Mann matrices $\hat{\sigma}_j$ of a $N$-state system \cite{NLevelBloch}. For a two-state system, the matrices $\hat{\sigma}_j$ for $j=x,y,z$ are the Pauli matrices, and $\qq = \{ x, y, z\}$ is a vector in Bloch sphere coordinates. The quantum state trajectory can be computed with an update equation of the form, $\qq_{k+1} = \bm{\mathcal{E}}[\bm{q}_k,r_k]$, taking into account the measurement back-action from the measurement readout $r_k$, and also considering the unitary evolution from the measured system's Hamiltonian. 

Since the distribution of the measurement readout only depends on the quantum state right before the measurement in this Markovian approach, we can then write the joint probability density function (PDF) of all measurement outcomes and state trajectories ${\cal P}_{\zeta} \equiv P(\{ \bm{q}_k \}_1^n,\{ r_k \}_0^{n-1}|\, \qq_0, \zeta)$ given an initial state $\qq_0$ and a set of other constraints $\zeta$ as,
\begin{align}\label{eq-mainpdf}
{\cal P}_{\zeta} = B_{\zeta}
\prod_{k=0}^{n-1}P(\bm{q}_{k+1}|\bm{q}_k,r_k)\, P(r_k|\bm{q}_k).
\end{align}
This is a time step product of $P(r_k|\qq_k)$, the conditional probability distribution for the measurement outcome $r_k$ given the system state before the measurement $\qq_k$, and $P(\qq_{k+1}| \qq_k, r_k)= \delta^d (\qq_{k+1}- \bm{\mathcal{E}}[\bm{q}_k,r_k])$, the (deterministic) conditional probability distribution for the quantum state after the measurement, given the state at the previous time step and the measurement readout. The prefactor $B_{\zeta} = B_{\zeta}[\{ \qq_k\}, \{ r_k\}]$ in Eq.~\eqref{eq-mainpdf} is a function of quantum states and readouts at any times, accounting for constraints used in selecting a sub-ensemble of the quantum trajectories, such as initial-state and final-state conditions.

The benefit of having the joint PDF Eq.~\eqref{eq-mainpdf} is that it contains all the statistical information about the system's evolution under measurement, and it allows us to selectively work with sub-ensembles of quantum trajectories simply by adding constraints (or conditions) to the joint distribution. Statistical moments can be computed from this joint PDF by integrating over its variables. For example, an expected value of an arbitrary functional ${\cal A} = {\cal A}[\{\qq_k\}, \{ r_k \}]$ is given by $\la {\cal A} \ra_{\zeta} = \int \!\!\dd[\qq_k]_1^n \dd[r_k]_0^{n-1} {\cal P}_{\zeta} {\cal A}$, where we define a notation for the multiple integral, $\int\!\! \dd [\qq_k]_{1}^{n} \equiv \int\!\! \dd \qq_{1} \cdots \dd \qq_{n}$. Direct integration of these quantities using the joint PDF as in Eq.~\eqref{eq-mainpdf}, however, can be a challenging task even for a simple qubit measurement problem. As such, we are motivated to write the joint PDF in a path integral form with an action (or exponent), so we can perform the integration using techniques developed in quantum theory such as a diagrammatic perturbation theory.
 
The path integral representation of the joint PDF in Eq.~\eqref{eq-mainpdf} can be attained by writing the delta functions for the state update $ \delta^d (\qq_{k+1}- \bm{\mathcal{E}}[\bm{q}_k,r_k])$ for $k=0$ to $n-1$ in the Fourier integral form, i.e.,  $\delta(q) = (1/2 \pi i) \int_{-i \infty}^{i \infty}  e^{-p q}\,\dd p$ for each $k$ and each component of the vector $\qq$, and then rewrite other terms in exponential forms. The conjugate variables for those delta functions are denoted by $\pp_k$ for $k=0$ to $n-1$. We refer to Ref.~\cite{Chantasri2013} and its Appendixes for a thorough discussion about this transformation and the construction of the path integral. As a result, the joint PDF is then given in a path integral form,
\begin{align}\label{eq-jointpdforig}
{\cal P}_{\zeta}= {\cal N}\!\!\int \!\!\dd [ \pp_k]_{0}^{n-1} \exp({\cal S}),
\end{align}
where the integrals are over all possible paths of $\{ \pp_k \}_0^{n-1}$ and the action ${\cal S}$ is defined as,
\begin{align}\label{eq-actiongenS}
{\cal S} = {\cal B}_{\zeta} + \sum_{k=0}^{n-1} \bigg\{\!- \bm{p}_k\! \cdot (\bm{q}_{k+1}\!-\bm{\mathcal{E}}[\bm{q}_k,r_k]) + \ln P(r_k | \bm{q}_k) \bigg\}.
\end{align}
We note that ${\cal B}_{\zeta}$ is an additional term determined by the formation of $B_{\zeta}$ in Eq.~\eqref{eq-mainpdf}, and ${\cal N}$ is a prefactor absorbing normalization constants.

For an example, we consider a sub-ensemble of trajectories that obey conditions on the initial and final states. The theoretical analysis of this kind of constraint is presented in Ref.~\cite{Chantasri2013}. The initial state is fixed at $\qq_0 = \qq_I$ and the final state is at $\qq_n = \qq_F$. This leads to the constraint term $B_{\zeta} = B_{\qq_I, \qq_F}= \delta^d(\qq_0 - \qq_I)\delta^d(\qq_n - \qq_F)$ in Eq.~\eqref{eq-mainpdf} and the preselected and postselected joint PDF, ${\cal P}_{\qq_I,\qq_F} = P(\{\qq_k\}_0^n, \{ r_k \}_0^{n-1} , \qq_F | \qq_I)$, which is written in a path integral form as,
\begin{align}\label{eq-jointpdfpps}
{\cal P}_{\qq_I,\qq_F} = {\cal N}\!\!\int \!\!\dd [ \pp_k]_{-1}^{n} \exp({\cal S}_{\qq_I, \qq_F}),
\end{align}
where the path integral's action is,
\begin{align}\label{eq-actiongenS-edit}
\nn{\cal S}_{\qq_I,\qq_F} = &-\pp_{-1}\cdot(\qq_0-\qq_I) - \pp_n\cdot(\qq_n-\qq_F)\\
+ &\sum_{k=0}^{n-1} \bigg\{\!- \bm{p}_k\! \cdot (\bm{q}_{k+1}\!-\bm{\mathcal{E}}[\bm{q}_k,r_k]) + \ln P(r_k | \bm{q}_k) \bigg\}.
\end{align}
We note that, in Eq.~\eqref{eq-jointpdfpps} and \eqref{eq-actiongenS-edit}, two additional conjugate variables, $\pp_{-1}$ and $\pp_n$, are introduced, because we have written the delta functions in $B_{\qq_I, \qq_F}$, one for the initial state and another for the final state, in the Fourier integral form. 

We can investigate the path integral's largest contribution by solving for its action's extrema. Taking the variation of the action Eq.~\eqref{eq-actiongenS-edit} over all the variables and setting it to zero, we obtain a set of difference equations,
\begin{subequations}\label{eq-diffeqs}
\begin{align}
 -\bm{q}_{k+1} +\bm{\mathcal{E}}[\bm{q}_k,r_k]&=0,\\
 \label{eq-backwardp} -\bm{p}_{k-1} + \frac{\partial}{\partial \bm{q}_k}\big\{\bm{p}_k\cdot \bm{\mathcal{E}}[\bm{q}_k,r_k] + \ln P(r_k | \bm{q}_k)\big\}&=0, \\
 \frac{\partial}{\partial r_k}\big\{ \bm{p}_k \cdot \bm{\mathcal{E}}[\bm{q}_k,r_k] + \ln P(r_k | \bm{q}_k)\big\}&=0.
\end{align}
\end{subequations}
The first, second and third equations are from taking derivatives over the conjugate variables $\pp_k$ from $k= 0$ to $n-1$, over the state variables $\qq_k$ from $k=1$ to $n-1$, and over the measurement readout variables $r_k$ from $k=0$ to $n-1$, respectively. The derivative over the final state variable $\qq_n$ gives a trivial equation $-\pp_n - \pp_{n-1}=0$, whereas the derivatives of the action over $\pp_{-1}$ and $\pp_n$ force the boundary conditions $\qq_0 = \qq_I$ and $\qq_n = \qq_F$ on the solutions of Eqs.~\eqref{eq-diffeqs}. We note that, in the case of no final boundary condition on the quantum state (i.e., $\pp_n = 0$ or $B_{\zeta} = B_{\qq_I} = \delta^d(\qq_0-\qq_I)$), the derivative of the action over $\qq_n$ instead gives $\pp_{n-1} = 0$, a final value of the conjugate variable. 

Interestingly, this extremization of the action in Eqs.~\eqref{eq-diffeqs} reproduces the Lagrange multiplier method, an optimization strategy that accommodates constraints. In our case, the optimized function is the last term of Eq.~\eqref{eq-actiongenS-edit}, $\sum_{k=0}^{n-1}\ln P(r_k | \bm{q}_k)$, subject to the constraints $\bm{q}_{k+1}=\bm{\mathcal{E}}[\bm{q}_k,r_k]$ for $k=0,...,n-1$ (enforcing the deterministic state update) with the Lagrange multipliers $ \bm{p}_{0}, ... , \bm{p}_{n-1}$, and the boundary constraints $\qq_0 = \qq_I$ and $\qq_n = \qq_F$ with the Lagrange multipliers $\pp_{-1}$ and $\pp_n$. Therefore, a solution of the difference equations Eqs.~\eqref{eq-diffeqs} is a path that optimizes the log-likelihood of a quantum trajectory, $\sum_{k=0}^{n-1}\ln P(r_k | \bm{q}_k)$, subject to the indicated constraints. The optimal path can be a local maximum, a local minimum, or a saddle point in the constrained probability space. For the optimal path that represents the local maximum, we call it the most likely path or the most probable path.

The optimal path, a solution of the difference equations Eq.~\eqref{eq-diffeqs} and their boundary conditions, can be approximated by taking a time-continuous limit $\delta t \rightarrow 0$, changing the difference equations to a set of ordinary differential equations, which can then be solved analytically or numerically. This approximation is applicable because its solutions, the optimal readouts and the optimal quantum paths, are smooth functions of time. In the case of the optimal paths that maximize the log-likelihood of preselected and postselected quantum trajectories (the most likely paths between two quantum states), these solutions have been experimentally verified with a superconducting transmon qubit \cite{Weber2014}.

So far, we have presented the joint PDF and the path integrals in the time-discrete form. For a more compact representation, we introduce a time-continuous version of the discrete stochastic path integral, assuming that the system is not intrinsically discrete and a well-defined diffusive limit exists. The joint PDF, for example in Eq.~\eqref{eq-jointpdforig}, is written as,
\begin{align}\label{eq-contpdf}
{\cal P}_{\zeta} = {\cal N}\!\!\int \!\!\dd [ \pp_k] \, \exp({\cal S}) \overset{\delta t \rightarrow 0}{=} {\cal N}\!\! \int \!\!\! {\cal D}\pp\, \exp({\cal S}),
\end{align}
where we have defined a notation for functional integrals, e.g., $\int \!\! {\cal D}\pp \equiv \lim_{\delta t \rightarrow 0} \int \!\! \dd [ \pp_k]$. The action in Eq.~\eqref{eq-actiongenS} is given by,
\begin{align}\label{eq-contaction}
{\cal S}=& \,\,{\cal B}_{\zeta} + \int_0^T \!\!\! \dd t \big\{\!\!- \pp\cdot ({\dot \qq} - {\bm {\mathcal L}}[\qq , r]) + {\cal F}[\qq,r]\big\},
\end{align}
where we have introduced ${\dot \qq}\,\dd t = {\bm {\mathcal L}}[\qq , r]\dd t$ as the time-continuous version of the state update equation $\qq_{k+1}=\bm{\mathcal E}[\qq_k,r_k]$, and defined ${\cal F}[\qq,r]$ as the linear-order expansion in time of the log-probability, i.e., $P(r_k |\qq_k) \propto \exp\{\delta t {\cal F}[\qq_k,r_k]+O(\delta t^2)\}$. A proportionality factor of the latter is absorbed into the normalization factor ${\cal N}$. The time dependence of the variables in Eq.~\eqref{eq-contpdf} and \eqref{eq-contaction} is suppressed for simplicity (e.g., $\qq \equiv \qq(t)$).

In the continuous version of the action, the state update equation ${\dot \qq} = {\bm {\mathcal L}}[\qq , r]$ can be derived from the first order expansion in $\delta t$ of its discrete form $\qq_{k+1}=\bm{\mathcal E}[\qq_k,r_k]$ using the exact readout probability distribution $P(r_k|\qq_k)$. As an alternative, one can consider using a stochastic master equation of the quantum state, where an ideal white noise $\xi$ is introduced with its single Gaussian probability distribution. The latter substitution is valid in an idealized noise limit, which we discuss in more detail in section \ref{subsec-nonqnd} (where It\^{o} stochastic equations are chosen in the diagrammatic perturbation approach) and in Appendix~\ref{app-whitenoise}. The choice of the stochastic state equations needs to be consistent with the readout (or noise) probability distribution used in obtaining the functional ${\cal F}[\qq,r]$ in Eq.~\eqref{eq-contaction}.

A careful analysis is needed when considering the extremization of the action as described in Eqs.~\eqref{eq-diffeqs}, which, in the continuum limit, changes to a set of ordinary differential equations. We note that one can derive this same set of differential equations directly from extremizing the continuous-version action in Eq.~\eqref{eq-contaction}, if the state update equation ${\dot \qq} = {\bm {\mathcal L}}[\qq , r]$ is obtained from the first order expansion in $\delta t$ of its discrete form (this approach to the derivation is presented in Ref.~\cite{Chantasri2013}). For the action constructed using the It\^{o} stochastic equations, its extremization does not give a correct set of differential equations describing the optimal paths. However, a stochastic path integral formulated using the It\^{o} equations is more advantageous when computing functional integrals, which is presented in section~\ref{sec-purbexpand}.

\section{Theoretical model: qubit measurement}\label{sec-qubit}
So far we have presented the formalism in the general context, the quantum measurement with measurement readouts $\{ r_k \}$ and corresponding quantum states $\{ \qq_k \}$. In order to show examples and demonstrate how to compute interesting quantities using the path integral formalism, we choose a particular theoretical model, a qubit continuously measured by an apparatus assuming a weakly responding (or coupled) detector. This theoretical setup has long been a subject of interest in theories and experiments. Some of the physical realizations of this system available with current technology are as follows: a single electron in a double quantum dot capacitively coupled to a quantum point contact detector \cite{Korotkov1999,Korotkov2001,Sun2001,Goan2001,Nori2009,Nori2009-2}, a superconducting qubit dispersively coupled to a microwave waveguide cavity \cite{Gambetta2008,Paik2011,Kater2013}, and quantum optics experiments such as a two-level atom monitored with homodyne optical measurement \cite{Wiseman1993-2}.

In most of these experimental setups, one can characterize the qubit evolution as governed by the measurement back-action, the qubit unitary evolution, and extra dephasing due to loss of information or added noise. We write a time-discrete change for a qubit density matrix $\rho$ as ${\rho}_{k+1}={\cal O}_{\gamma}\,{\cal U}_{\delta t}\,{\cal M}_{r_k}[\rho_k]$, where we define a measurement operation ${\cal M}_{r_k}[\rho] ={\op M}_{r_k} \rho {\op M}_{r_k}^{\dagger}/\Tr{{\op M}_{r_k} \rho {\op M}_{r_k}^{\dagger}}$, a unitary operation ${\cal U}_{\delta t}[\rho] = e^{-i {\op H} \delta t} \rho e^{i {\op H} \delta t}$, and an extra dephasing operation ${\cal O}_{\gamma}[\rho]$ (we set $\hbar = 1$). 

The measurement back-action on a qubit state, considering the diffusive measurement as in Refs.~\cite{Korotkov1999,Korotkov2001,Chantasri2013}, is described by a measurement operator ${\op M}_{r_k} = (\delta t/2 \pi \tau_m)^{1/4} \exp[-\delta t(r_k-{\op \sigma}_z)^2/4 \tau_m]$, where $\tau_m$ is a characteristic measurement time taken to separate the two Gaussian distributions by two standard deviations. The measurement readout distribution is computed from the trace of the measurement operator and the qubit state,
\begin{align}\label{eq-probr}
\nn P(r_k | \rho_k)\, =&\,\,\Tr{{\op M}_{r_k} \rho_k {\op M}_{r_k}^{\dagger}},\\
\nn=&\, \left(\frac{\delta t}{2 \pi \tau_m}\right)^{\frac{1}{2}}e^{-\frac{\delta t}{2 \tau_m}(r_k-1)^2}(1+z_k)/2 \\ 
 &+\left(\frac{\delta t}{2 \pi \tau_m}\right)^{\frac{1}{2}} e^{-\frac{\delta t}{2 \tau_m}(r_k+1)^2}(1-z_k)/2,
\end{align}
with the Bloch sphere coordinates of the qubit state given by $\{ x_k , y_k, z_k \}$. 

We denote a qubit Hamiltonian by ${\op H} = (\epsilon/2) {\op \sigma}_z + (-\Delta/2){\op \sigma}_x$ characterizing the unitary evolution during the measurement, and we define the dephasing operation ${\cal O}_{\gamma}[\rho]$ accounting for additional dephasing on the system, such as detection inefficiency and dephasing due to the environment. We model the dephasing operation as an extra dephasing rate $\gamma$ on the off-diagonal elements of the qubit density matrix $\rho$. By expanding the state update equation ${\rho}_{k+1}={\cal O}_{\gamma}\,{\cal U}_{\delta t}\,{\cal M}_{r_k}[\rho_k]$ to first order in $\delta t$, and taking a continuum limit $\delta t \rightarrow 0$, we obtain a set of differential equations,
\begin{subequations}\label{eq-contstate}
\begin{align}
\dot{x} &=- \gamma \, x - \epsilon \, y - x \, z \, r /\tau_m,\\
\dot{y} &=-\gamma \, y +\epsilon \, x + \Delta \, z - y\, z \, r /\tau_m, \\
\dot{z} &= - \Delta \,y + (1-z^2)\,r/\tau_m,
\end{align}
\end{subequations}
which turns out to be analogous to the stochastic master equation in Stratonovich interpretation as mentioned in Ref.~\cite{Korotkov1999,Korotkov2001}. 

Using the state update equations in Eqs.~\eqref{eq-contstate} and the readout distribution in Eq.~\eqref{eq-probr}, we have the qubit action in continuous form (as in Eq.~\eqref{eq-contaction}),
\begin{align}\label{eq-contactionqubit}
\nn{\cal S}_{qb} = {\cal B}_{\zeta} + \int_0^T \!\!\!\dd t\,\big\{\! &- p_x (\dot{x} + \gamma \,x + \epsilon \, y + x\,z\,r/\tau_m)\\
\nn& - p_y(\dot{y}+\gamma \,y - \epsilon \, x -\Delta \, z + y\, z\, r/\tau_m)\\
\nn&-p_z (\dot{z}+ \Delta \, y- (1-z^2)\,r/\tau_m)\\
& - (r^2 - 2 \,r \, z + 1)/2 \tau_m\big\},
\end{align}
where we have approximated the logarithm of the readout distribution as $\ln P(r_k | z_k ) \approx - (\delta t/2 \tau_m) (r_k^2 - 2 r_k z_k +1) + (1/2) \ln (\delta t/2 \pi \tau_m) + O(\delta t^2)$, omitting the second term and its higher order expansion because they can be absorbed into the prefactor ${\cal N}$. We note that the extremization of this action and its most likely paths with fixed initial and final states are presented in more detail in Ref.~\cite{Chantasri2013}.

In the remainder of this paper, we will focus on applying the path integral formalism to this particular qubit measurement system, though in different regimes of the Hamiltonian parameters. For example, in the next section, we study the ``plain'' qubit measurement where we assume that there is no qubit Hamiltonian, i.e., $\epsilon = \Delta = 0$, and the qubit's evolution comes only from the measurement and the dephasing mechanism.

\section{Expansion around the optimal path in the plain measurement case}\label{sec-qnd}
In this section, we present examples as to how we can use the path integral and its optimal path to compute statistical moments of the measured qubit trajectories. As it turns out, analytic solutions are possible for the case of quantum non-demolition measurement ($\Delta =0$, where the system Hamiltonian commutes with total system-detector Hamiltonian) with a constraint on the initial and final states of the quantum trajectories. In this case, the path integral can be written only in the $z$-coordinate, i.e., ${\bm q} = z$, because the evolution of $z$ is independent of the other two coordinates, $x$ and $y$. The $x$ and $y$ degrees of freedom can be found directly from $r$ and $z$, so they will be dropped from the discussion until section~\ref{sec-purbexpand}. 

In the following subsections, we consider the plain measurement case when $\epsilon = \Delta = 0$ for simplicity. We show how to compute the path integral using the preselected and postselected joint PDF, $P(\{ z_k \}_0^{n},\{ r_k \}_0^{n-1},z_F | z_I)$. We first integrate this joint PDF over all intermediate variables to get the probability density function $P(z_F|z_I)$ of arriving at the final state $z_F$ after time $T$, given the initial state $z_I$. Then, we show how to generalize the integration procedures to compute statistical quantities such as averages, variances, and correlation functions of the qubit trajectories in this simplified case. We note here that even though we present the derivation in the plain measurement case, the result should still be valid for the case when $\epsilon \ne 0$. In the latter case, the evolution of the $x$ and $y$ coordinates of the qubit will be deterministically oscillating with the frequency $\epsilon$, without affecting the evolution of the $z$ coordinate.

\subsection{Probability density function of a final state given an initial state}
In order to compute the probability distribution of the final qubit state given the initial state separated by time $T$, we start with the joint PDF ${\cal P}_{z_I,z_F}=P(\{ z_k \}_0^{n},\{ r_k \}_0^{n-1},z_F | z_I)$ of the qubit trajectories $\{ z_k \}_0^n$ and measurement outcomes $\{ r_k \}_0^{n-1}$ with fixed boundary states, and then integrate over all variables except the final state $z_F$,
\begin{align}\label{eq-intprobzf}
P(z_F| z_I) = \!\!\int\!\!\dd [z_k]_0^{n}\dd [r_k]_0^{n-1} {\cal P}_{z_I,z_F},
\end{align}
where, as before, we have defined the multi-variable integral, e.g., $\int\!\! \dd [z_k]_{0}^{n} = \int\!\! \dd z_{0} \cdots \dd z_{n}$. 

From the discussion in section~\ref{sec-mainSPI}, the joint PDF is given by a product over the sliced time variables, ${\cal P}_{z_I,z_F} = \delta(z_0 - z_I) \delta(z_n-z_F)\prod_{k=0}^{n-1}P(z_{k+1}|z_k,r_k)\, P(r_k|z_k)$. The term describing the state update equation is a delta function,
\begin{align}\label{eq-deltaz}
P(z_{k+1}|z_k,r_k) = \delta\bigg(z_{k+1} - \frac{z_k \cosh \frac{r_k \delta t}{\tau_m}  + \sinh \frac{r_k \delta t}{\tau_m} }{\cosh \frac{r_k \delta t}{\tau_m} + z_k \sinh \frac{r_k \delta t}{\tau_m} }\bigg),
\end{align}
with its argument derived from the $z$-coordinate of the density matrix equation $\rho_{k+1} = {\cal O}_{\gamma} {\cal U}_{\delta t} {\cal M}_{r_k}[\rho_k]$, whereas, the probability distribution for the readout Eq.~\eqref{eq-probr} is expressed in this form,
\begin{align}\label{eq-dtprobr}
 P(r_k | z_k)\, \approx & \left(\frac{\delta t}{2 \pi \tau_m}\right)^{\frac{1}{2}}\!\!\exp\left\{  \frac{-\delta t}{2 \tau_m}  \big( r_k^2 - 2 r_k z_k + 1)\right\},
\end{align}
neglecting higher orders in $\delta t$. Substituting these two expressions, Eqs.~\eqref{eq-deltaz} and \eqref{eq-dtprobr}, into the joint PDF, we get
\begin{align}
\nonumber{\cal P}_{z_I,z_F} \!=\,&\,\delta(z_n-z_F)\delta(z_0-z_I)\left(\frac{\delta t}{2 \pi \tau_m}\right)^{\frac{n}{2}} \left\{\prod_{k=0}^{n-1}\delta(z_{k+1} \cdots )\right\}
\\& \times \exp\bigg\{-\sum_{k=0}^{n-1} \frac{\delta t}{2\tau_m}(r_k^2 - 2 r_k z_k +1)\bigg\},
\end{align}
where we have used $\delta(z_{k+1} \cdots)$ to represent the delta function in Eq.~\eqref{eq-deltaz}. 

We then integrate out the measurement readouts by transforming the delta function of $z$ into a delta function in the readout variable, $\delta(z_{k+1} \cdots)= \delta(r_k - \frac{\tau_m}{\delta t}\tanh^{-1}z_{k+1}+\frac{\tau_m}{\delta t}\tanh^{-1}z_k)\tau_m/(1-z_{k+1})^2\delta t$, using the transformation relation $\delta[x-f(y)] = \sum_i \delta[y-y_i]/|\partial_y f(y)|_{y_i=f^{-1}(x)}$, where the sum extends over all solutions of $y_i = f^{-1}(x)$. After integrating over the measurement readout and also over the boundary state $z_0$, and $z_n$, the joint PDF is transformed to a new joint PDF, a function of only the intermediate $z$-variables,
\begin{align}\label{eq-probzz}
 P( \{ z_k \}_1^{n-1}, z_F | z_I) = \left(\frac{\tau_m}{2 \pi \delta t}\right)^{\frac{n}{2}} \left(\prod_{k=0}^{n-1} \frac{1}{1-z_{k+1}^2} \right)\exp( {\cal S}),
\end{align}
where the action ${\cal S}$ is,
\begin{align}\label{eq-probzz2}
{\cal S} = - \delta t \sum_{k=0}^{n-1} \bigg \{\frac{ \tau_m}{2}\frac{ (u_{k+1}-u_k)^2 }{ \delta t^2} -\tanh u_k \frac{(u_{k+1}-u_k)}{\delta t}+ \frac{1}{2 \tau_m}\bigg\}.
\end{align}
We note that we have simplified the above equation by defining a new variable $u_k \equiv \tanh^{-1} z_k$ to write the joint PDF in a compact form. This joint PDF of the $z$-variable is one of the main results shown in this paper. It is also important to point out that since we have already integrated over the boundary state variable $z_0$ and $z_n$, the delta functions for the boundary states have applied to the states $z_0 = z_I$ and $z_n = z_F$ in Eqs.~\eqref{eq-probzz} and \eqref{eq-probzz2}.

The next step is to perform the integration over the intermediate variables $z_k$ for $k=1,...,n-1$. These integrals can be done easily by expanding $z_k$ around the optimal solution denoted by ${\bar u}_k$. We substitute $u_k = {\bar u}_k +  \eta_k$ for all $k$'s in the action, and perform integration by parts with the vanishing boundaries, $\eta_0=\eta_n=0$. This substitution exactly simplifies to ${\cal S}[u] = {\cal S}[{\bar u}] - \frac{\tau_m}{2 \delta t} \sum_{k=0}^{n-1}(\eta_{k+1}-\eta_k)^2$ (see Appendix~\ref{app-actionpure} for more detail). We then change the integral measures to $\dd \eta_k \equiv \dd z_k/(1-z_k^2)$ for $k = 1,...,n-1$ and write the probability distribution in Eq.~\eqref{eq-probzz} in terms of $\eta$-variables,
\begin{align}\label{eq-derivep}\nn
\nn P( \{ \eta_k \}_1^{n-1}, z_F& | z_I) =   \left(\frac{\tau_m}{2 \pi \delta t}\right)^{\frac{n}{2}}\frac{ \exp({\cal S}\,[\bar{u}])}{1-z_{F}^2} \\
&\times  \exp\left\{-  \frac{ \tau_m}{2 \delta t}\sum_{k=0}^{n-1}(\eta_{k+1}-\eta_k)^2\right\},
\end{align}
where the optimal path ${\bar u}_k$ is a solution of the extremization of the action, which gives $\sum_{k=0}^{n-1}({\bar u}_{k+1}-2 {\bar u}_k+{\bar u}_{k-1})\delta t^{-1} = 0$. The solution is ${\bar u}_k = c_1 t_k + c_2$ with the two constants of integration $c_1 = \frac{1}{T}(\tanh^{-1}z_F - \tanh^{-1}z_I)$ and $c_2 = \tanh^{-1} z_I$. The optimal solution (the most likely path) is given by,
\begin{align}\label{eq-ubar}
{\bar u}_k = \frac{t_k}{T}(\tanh^{-1}z_F - \tanh^{-1}z_I) + \tanh^{-1} z_I,
\end{align}
where we wrote $t_k = k \delta t$. We note that in the time-continuum limit, this extremization is equivalent to the vanishing functional derivative of the action, $\delta {\cal S}[u]/\delta u(t) = \tau_m \ddot{u} = 0$, which is similar to the Euler-Lagrange equation for the classical trajectory of a free particle in the $u$ coordinate transformation.

The integrals in Eq.~\eqref{eq-intprobzf}, with the probability distribution in $\eta$-variable Eq.~\eqref{eq-derivep}, is now in this form,
\begin{align}\label{eq-intprobzfeta}
P(z_F|z_I) = \int \!\!\dd[\eta_k ]_1^{n-1} P( \{ \eta_k \}_1^{n-1}, z_F | z_I).
\end{align}
Fortunately, they are Gaussian integrals in multiple dimensions which can be calculated from the matrix integral, $\int \!\dd {\bm \eta} \exp(-\frac{1}{2}{\bm \eta}^T\cdot {\bm M} \cdot {\bm \eta}) = (2 \pi)^{\frac{n-1}{2}}/(\text{Det}{\bm M})^{1/2}$, where, in our case, $\int \! \dd{\bm \eta} = \int \!\! \dd[\eta_k]_1^{n-1}= \int \!\! \dd \eta_1\cdots \dd \eta_{n-1}$, and the matrix ${\bm M}$ and the vector ${\bm \eta}$ are given by,
\begin{align}\label{eq-matrixB}
{\bm M}=\frac{\tau_m}{ \delta t} \left ( \begin{matrix} 2 & -1 & 0 & \cdots \\ -1 & 2 & -1 & \cdots \\ 0 & -1 & 2 & \cdots \\ \vdots & \vdots & \vdots & \ddots \end{matrix} \right), \text{ and}\quad {\bm \eta} = \left( \begin{matrix} \eta_1 \\ \eta_2 \\ \vdots \\ \eta_{n-1} \end{matrix} \right),
\end{align}
noting that $\text{Det}{\bm M} = n (\tau_m/\delta t)^{n-1}$. 

After substituting the optimal solution ${\bar u}_k$ into ${\cal S}[\bar{u}]$ and performing the integrals over $\eta_1,...,\eta_{n-1}$ in Eq.~\eqref{eq-intprobzfeta}, we obtain the distribution of the final state fixing the initial state and the duration of time $t_n = T$,
\begin{align}\label{eq-probzf}
P( z_F| z_I) = \frac{\sqrt{\frac{\tau_m}{2 \pi T}}}{(1-z_F^2)} \exp\left\{ \frac{-T}{2 \tau_m}(\bar{r}^2+1) +\frac{1}{2} \ln \left(\frac{1-z_I^2}{1-z_F^2}\right)\right\},
\end{align}
where we have defined $\bar{r} = \frac{\tau_m}{T} (\tanh^{-1} z_F - \tanh^{-1} z_I)$ as the optimal measurement readout. This solution is exactly the same as what we would get from the change of variables of the probability distribution, $ P(z_F | z_I) \dd z_F = P(r_{\rm tot} | z_I) \dd r_{\rm tot} $, where $r_{\rm tot} =(1/n) \sum_{k=0}^{n-1} r_k = (\tau_m/T)(\tanh^{-1}z_F - \tanh^{-1} z_I)$ is a time-average measurement readout. The derivation of the latter distribution is presented in the methods section of Ref.~\cite{Weber2014}.

\subsection{Means and variances of qubit trajectories in the plain measurement case}
We have derived the probability distribution for the measured qubit given the fixed initial and final states, as shown in Eq.~\eqref{eq-probzz} and \eqref{eq-derivep}. In this subsection, we consider computing the qubit trajectory's statistical quantities that can be written in this expectation form,
\begin{align}\label{eq-expectva}
\nn  _{z_F}\langle {\cal A} \rangle_{z_I} \equiv & \int {\rm d}[z_k]_1^{n-1}{\cal A}\,\, \frac{P( \{ z_k \}_1^{n-1},z_F | z_I)}{P(z_F|z_I)},\\
=&\frac{\sqrt{\text{Det}{\bm M}}}{(2 \pi)^{\frac{n-1}{2}}} \int {\rm d}[\eta_k]_1^{n-1}{\cal A}\,\, e^{-\frac{1}{2}{\bm \eta}^T\cdot {\bm M} \cdot {\bm \eta}},
\end{align}
where ${\cal A}$ is an arbitrary functional of $z_j$ at any time $t_j$ for $j = 1,...,n-1$ and the matrix ${\bm M}$ is the same as defined in Eq.~\eqref{eq-matrixB}. Note that we have used the notation $_{z_F}\la \cdots \ra_{z_I}$ for a statistical average conditioned on fixed initial and final states, $z_I$ and $z_F$. Since the probability distribution in the $\eta$-variables is Gaussian, it is preferable to write $z$ in terms of $\eta$, replacing $z$ with $z = \tanh(\bar{u} + \eta)$, and then perform the integration. For example, a conditional average of $z_j$ at time $t_j$ is given by,
\begin{align}\label{eq-expandz}
\nn _{z_F}\langle z_j   \rangle_{z_I} =& _{z_F}\langle \tanh(\bar{u}_j+\eta_j)\rangle_{z_I}\\
\nn =&\tanh \bar{u}_j +_{z_F}\!\!\langle \eta_j  \rangle_{z_I}\tanh' \bar{u}_j \\
& + \frac{_{z_F}\langle \eta_j^2 \rangle_{z_I} }{2!}\tanh''\bar{u}_j + { O}(\eta_j^3),
\end{align}
expanding to second order in $\eta$, where the primes indicate derivatives over the $u$-variable. In the second line, we still use the bracket $_{z_F}\la \cdots \ra_{z_I}$ for the conditional average of the variable $\eta$, even if its boundary values are shifted to $\eta_0 = \eta_n = 0$.

The expectation values in terms of $\eta$ can be computed using the definition of moments in the second line of Eq.~\eqref{eq-expectva}. Because it is a multi-dimensional Gaussian integral, we find the result from Wick's theorem,
\begin{align}\label{eq-wick}
 _{z_F}\langle \eta_{j_1} \eta_{j_2} \cdots  \eta_{j_{2m}} \rangle_{z_I} 
& =\!\!\!\!\! \sum_{\substack{\text{all possible} \\ \text{pairings of} \\ \{j_1, j_2, \dots, j_{2m}\}}} \!\!\!\!\! M^{-1}_{j_{k_1},j_{k_2}}\cdots \,\,M^{-1}_{j_{k_{2m-1}},j_{k_{2m}}},
\end{align}
where the left hand side is an expectation value of even numbers of $\eta$ at times $t_1, t_2, ..., t_{2m}$, and the right hand side is a product of $m$ elements of the inverse matrix ${\bm M}^{-1}$, summing over all possible pairings between the $\eta$'s on the left side. Any other statistical averages with odd numbers of $\eta$ will vanish. As an example, pairing four $\eta$-variables, $_{z_F}\langle \eta_j \eta_k \eta_l \eta_m \rangle_{z_I}$, yields the sum, $M^{-1}_{jk}M^{-1}_{lm} + M^{-1}_{jl}M^{-1}_{km} +M^{-1}_{jm}M^{-1}_{kl}$.

\begin{figure}
\includegraphics[width=8.5cm]{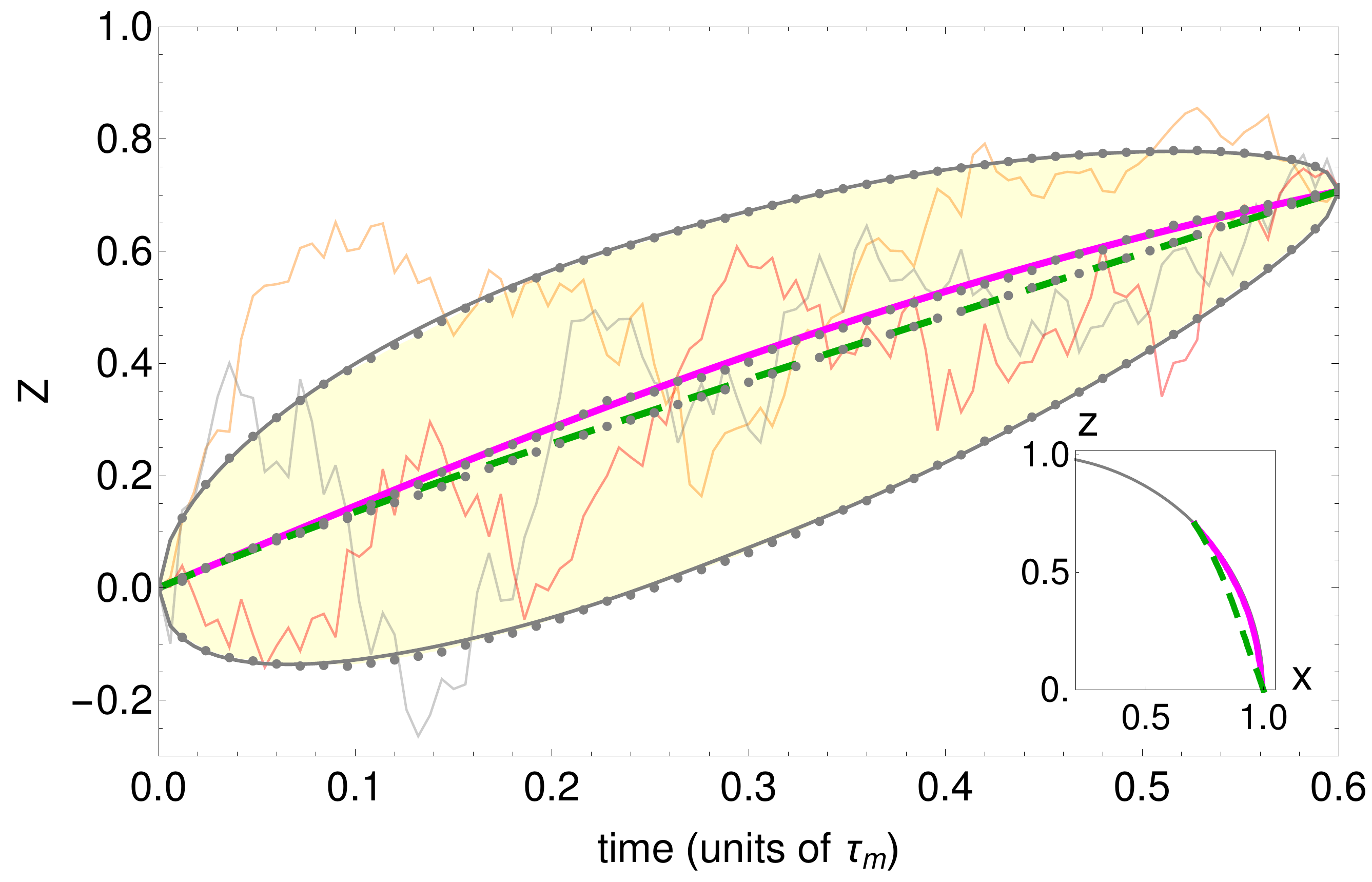}
\caption{(Color online) The comparison between the analytical solutions, Eq.~\eqref{eq-ubar},\eqref{eq-averagez},\eqref{eq-variancez}, and the numerical simulation, for the preselected and postselected average (green dashed curve), the most likely path (magenta curve), and the variance (the two gray thin curves on both sides of the average are the average $\pm$ standard deviation) of the measured qubit trajectories. The preselected state (initial state) and the postselected state (final state) are $z_I = 0$ and $z_F = \cos \pi/4$, respectively. The total time is $T = 0.6 \, \tau_m$. The numerical data, plotted as gray dots, are analyzed from $5\times 10^5$ trajectories computed with time steps of the size $\delta t  = 0.006 \, \tau_m^{-1}$, zero qubit Hamiltonian, and the final selection tolerance $z_F \pm 0.02$. The numerical data shows excellent agreement with the theoretical solutions. The three fluctuating curves are randomly chosen individual trajectories. The inset shows the average (dashed green) and the most likely path (magenta) on a Bloch sphere plotted in the $x$-$z$ plane.}
\label{fig-qnd}
\end{figure}

The elements of the inverse matrix ${\bm M}^{-1}$ are of a simple form, $M^{-1}_{jk} = M^{-1}_{kj}=(\delta t/\tau_m)j(n-k)/n$ for $k\ge j$ (the full matrix is presented in Appendix~\ref{app-invmatrix}). Knowing these matrix elements, we can then compute any expectation values of the type shown in Eq.~\eqref{eq-wick}, such as a two-time correlation,
\begin{align}\label{eq-correta}
_{z_F}\langle \eta_j \eta_k \rangle_{z_I} =M_{jk}^{-1} =  \frac{ t_j}{\tau_m}\left(1-\frac{t_k}{T}\right),
\end{align}
for $k \ge j$, and statistical moments of even orders of $\eta$,
\begin{align}\label{eq-etamo}
 _{z_F}\langle \eta_j^{2p} \rangle_{z_I} =& (2p-1)!!\left(\frac{ t_j}{\tau_m}\right)^p \left(1-\frac{t_j}{T}\right)^p,
\end{align}
where we have used the discrete time notation $t_j = j \delta t$ and $T = t_n = n \delta t $, and the double factorial prefactor $ (2p-1)!! = \frac{(2p)!}{2^p p!}$ for a positive integer $p$. This double factorial comes from the number of ways of pairing $2p$ identical variables.  

Substituting these quantities into the expectation value Eq.~\eqref{eq-expandz}, we then obtain the preselected and postselected average of $z_j$,
\begin{align}\label{eq-averagez}
 _{z_F}\langle z_j \rangle_{z_I} \approx \, &\tanh \bar{u}_j - \frac{t_j}{\tau_m}\left(1-\frac{t_j}{T}\right)\sech^2 {\bar u}_j \tanh {\bar u}_j,
\end{align}
keeping terms up to first order of $T/\tau_m$, which is equivalent to second order of $\eta$ because the conditional average of the second order of $\eta$ scales as $T/\tau_m$, i.e., $_{z_F}\langle \eta_j \eta_j \rangle_{z_I} \sim T/\tau_m$. This solution Eq.~\eqref{eq-averagez} is justified in a weak-coupling limit ($\tau_m \gg T$), however, the full solution to all orders can be computed and is presented in Appendix~\ref{app-fullqnd}. Using the same approximation, we can also compute another interesting quantity, the variance, keeping terms up to first order of $T/\tau_m$,
\begin{align}\label{eq-variancez}
\nn _{z_F}\langle \Delta z_j^2\rangle_{z_I} \equiv & \, _{z_F}\langle z_j^2 \rangle_{z_I} - \, \left( _{z_F}\langle z_j \rangle_{z_I}\right)^2\\
\approx &\, \frac{t}{\tau_m}\left(1-\frac{t}{T}\right) \sech^4 \bar{u}_j,
\end{align}
where its full solution to all orders in $T/\tau_m$ is also presented in Appendix~\ref{app-fullqnd}. Finally, we compute a two-time correlation in $z$-coordinate, keeping up to first order in $T/\tau_m$,
\begin{align}
_{z_F}\la z_j z_k \ra_{z_I} \approx \frac{t_j}{\tau_m}\left(1-\frac{t_k}{T}\right)\sech^2 {\bar u}_j \, \sech^2 {\bar u}_k,
\end{align}
for $ t_k \ge t_j$, which decreases linearly in $t_k$ for a fixed $t_j$.

We show in Figure~\ref{fig-qnd}, the average $ _{z_F}\langle z_j \rangle_{z_I}$, its variance, and the optimal path (the most likely path) ${\bar u}_j $, for the preselected and postselected trajectories, compared with data from a numerical simulation using the Monte Carlo method (see Appendix~\ref{app-numer} for more detail). We generate $5\times 10^5$ trajectories with a postselection time $T = 0.6 \, \tau_m$, showing an excellent agreement with the analytical solutions.

\section{Diagrammatic expansion theory}\label{sec-purbexpand}
We have shown in the previous section that the preselected and postselected moments (or expected values) for quantum trajectory variables, in the plain measurement case, can be computed by expanding the path integral around its optimal path. Since the action is Gaussian, however, the path integral approach can be useful in other cases as well, such as to compute the moments and expectation values of qubit trajectories when there is no final state condition (state postselection), or with non-zero qubit Hamiltonian. In this section, we present an alternative method using a diagrammatic perturbation expansion of the action, similar to how Feynman diagrams are used in computing path integrals in quantum field theory.

In the following, we start with a brief introduction to the perturbation method (a standard method used in quantum field theory) and show how one can compute expectation values of system variables from moment generating functionals. We will then go on to the examples, for the case of qubit trajectories with non-zero Hamiltonian and no state postselection. We note that even though most of the derivations are shown in time-continuous form for simplicity, the most straightforward derivations are in time-discretized version, and some of these are shown in Appendixes~\ref{app-sourceterm} and \ref{app-heaviside}.

\subsection{Brief introduction to the perturbation expansion}\label{sec-briefintro}
Let us suppose that a quantity of interest is an expectation value given in a path integral form,
\begin{align}\label{eq-expectA}
\langle {\cal A} \rangle \equiv \,\,{\cal N}\!\! \int  \!\!{\cal D}\bxx {\cal D}\bxt e^{{\cal S}}{\cal A},
\end{align}
where the quantity ${\cal A}$ is a functional of a system variable $\bxx$, the integral's action ${\cal S}$ is a functional of the system variable $\bxx$ and its conjugate variable $\bxt$, and ${\cal N}$ is a constant normalized factor. Although we are writing $\bxx$ and $\bxt$ as one-dimensional variables here, a generalization to arbitrary dimensional vectors is straightforward.

To compute the integral above, we write the action as a sum of two parts, ${\cal S} = {\cal S}_F + {\cal S}_I$, one being terms in the action that have bilinear forms, (e.g., $\sim \!\bxx\bxt$), which we call the \textit{free action},
\begin{align}\label{eq-bilinearS}
{\cal S}_F =\, - \!\!\int\!\! \dd t \dd t'\bxt(t) G^{-1}(t,t') \bxx(t'),
\end{align}
and call the rest of terms in the action the \textit{interaction action} ${\cal S}_I$. We then define a free generating functional ${\cal Z}_F[\bjj, \bjt]$, a path integral of the free action adding extra source terms with functions $\bjj$ and $\bjt$,
\begin{align} \label{eq-genfunZ} 
\nn {\cal Z}_F[J, \tilde{J}] =&\,\,{\cal N}\!\!\int\!\! \mathcal{D} \bxx \mathcal{D} \bxt e^{{\cal S}_F + \int\!\! \dd t \bxt(t) \bjt(t) + \int \!\!\dd t \bjj(t) \bxx(t)  }\\ 
= &\,\,\exp\left\{\int \!\!\dd t \dd t' \bjj(t){ G}(t,t')\bjt(t')   \right\},
\end{align}
where $G(t,t')$ is an inverse of $G^{-1}(t,t')$ in Eq.~\eqref{eq-bilinearS} satisfying $\int \!\!\dd t'' G^{-1}(t, t'')G(t'',t') = \delta (t-t')$, and we assume that the Gaussian integrals in the first line produce a factor ${\cal N}^{-1}$ that leaves no prefactor in the second line. Note that these integrals converge when $\bxt$ is purely imaginary. Moreover, if the bilinear terms are instead quadratic terms (e.g., $\sim X^2$), the free action will be of the form ${\cal S}_{F,{\rm quad}} = - \frac{1}{2}\int \!\!\dd t \dd t' \bxx(t) G^{-1}(t,t') \bxx(t')$, and there will be only one source term $\int \!\!\dd t X(t) J(t)$ leading to a different generating functional ${\cal Z}_{F,{\rm quad}}[\bjj,\bjt] = \exp\left\{\frac{1}{2}\int\!\!\dd t\dd t' \bjj(t) G(t,t') \bjj(t')\right\}$.

The generating functional Eq.~\eqref{eq-genfunZ} is then used to compute a \textit{free moment} defined as $\langle \cdots \rangle_F \equiv {\cal N}\! \int \! {\cal D}\bxx {\cal D}\bxt e^{{\cal S}_F}\cdots $. The free moment of the variable $\bxx(t)$ (or $\bxt(t)$) at time $t$ is simply a functional derivative of the generating functional over the variable $\bjj(t)$ (or $\bjt(t)$), taking both variables $J$ and $\bjt$ to be zero at the end. By looking at the second line of Eq.~(\ref{eq-genfunZ}), one can see that the simplest non-vanishing free moment is a two-point correlation function $\langle \bxx(t)\bxt(t')  \rangle_F = \frac{\delta}{\delta \bjj(t)}\frac{\delta}{\delta \bjt(t')}{\cal Z}_F[\bjj,\bjt]\big|_{\bjj = \bjt = 0} = G(t,t')$. Generalizing this to moments of multiple time points, we get
\begin{align}\nn
\langle \bxx&(t_{j_1})\bxx(t_{j_2}) \cdots \bxt(t_{j_{2m-1}}) \bxt(t_{j_{2m}}) \rangle_F\\
\nn=&\frac{\delta}{\delta \bjj(t_{j_1})}\frac{\delta}{\delta \bjj(t_{j_2})}\cdots \frac{\delta}{\delta \bjt(t_{j_{2m-1}})}\frac{\delta}{\delta \bjt(t_{j_{2m}})}\,{\cal Z}_F[\bjj,\bjt]\bigg|_{\bjj=\bjt=0},\\
\nn=&\sum_{\substack{\text{all pairings between}\\ \text{variables ${\tilde X}$ and $X$}}} G(t_{j_{k_1}}, t_{j_{k_2}})\, \cdots \,G(t_{j_{k_{2m-1}}}, t_{j_{k_{2m}}}),
\end{align}
where the summation is for all possible pairings between the variables $\bxx$'s and their conjugate $\bxt$'s. The number of propagators in the summation on the right is equal to the number of pairs presented in the expectation bracket on the left. From this, one can see that if there is at least one unpaired variable, the moment will vanish.

Using these definitions of the free moments, one can then compute a path integral as in Eq.~\eqref{eq-expectA} where ${\cal S} = {\cal S}_F + {\cal S}_I$ and the arbitrary functional ${\cal A}$ is a function of the physical variable $\bxx$,
\begin{align}\nn
\langle {\cal A}[\bxx] \rangle \equiv &\,\,{\cal N}\!\! \int  \!\!{\cal D}\bxx {\cal D}\bxt e^{{\cal S}_F+{\cal S}_I[\bxx,\bxt]}{\cal A}[\bxx],\\
\nn=&{\cal A}\left[\frac{\hat \delta}{\delta \bjj}\right]e^{{\cal S}_I[\frac{\hat\delta}{\delta \bjj},\frac{\hat\delta}{\delta \bjt}]}{\cal Z}_F[\bjj,\bjt]\bigg|_{\bjj=\bjt=0},\\
=& \langle {\cal A}[\bxx]e^{{\cal S}_I[\bxx,\bxt]}\rangle_F, \label{eq-expA}
\end{align}
where in the second line we wrote the interaction action ${\cal S}_I$ and the arbitrary functional ${\cal A}$ as operators acting on the free generating functional ${\cal Z}_F[\bjj,\bjt]$.

The perturbation expansion refers to the series expansion of the term $\exp\{{\cal S}_I[\bxx,\bxt]\}$ in Eq.~\eqref{eq-expA}. In some cases, infinite series can be summed over, giving an exact analytic result, although in many others, an approximation is needed in order to truncate the series. An approximation can be made when there is a small parameter appearing in any (or all) of the terms in the interaction action ${\cal S}_I$, where the expansion is straightforward, keeping terms up to any desired order of the small parameter. Another type of approximation, which we present here in more detail, is similar to a semiclassical expansion in quantum mechanics, keeping terms up to any order of a small parameter $\nu$ that appears as an inverse in front of the action, for example,
\begin{align}\label{eq-actionsemiclass}
{\cal S} = \frac{1}{\nu}\left\{ -\!\!\int\!\! \dd t \dd t'\bxt(t) G^{-1}(t,t') \bxx(t') +{\cal S}_I[\bxx,\bxt] \right\},
\end{align}
where we explicitly write the the free action in a bilinear form. To compute an expectation value of a functional ${\cal A}$ using this action, one needs to expand the exponential of the interaction action and then evaluate the free moments in terms of the propagators. The order of expansion is controlled by the parameter $\nu$. There is a factor of $1/\nu$ for every single term in the expansion of ${\cal S}_I$, and a factor of $\nu$ for every propagator that emerges. This is the basic idea behind the loop expansion in quantum theory, where $\nu$ plays the role of the $\hbar$ in the Feynman path integral. We will discuss more about the loop expansion in section~\ref{sec-smallnoiseapprox}.

This loop expansion based on the small parameter $\nu$, in our quantum measurement case, can be considered as a small noise expansion around a saddle point solution (a solution that extremizes the action). Moreover, as in the Feynman path integral, under the approximation of the small parameter $\nu$, a saddle point approximation can also be applied to estimate a path integral, using an expansion of the action around the saddle point solution. For our stochastic path integral, the saddle point approximation gives an estimation of total probability density for the joint probability distribution that the path integral represents.

\subsection{Examples in continuous quantum measurements}
Now we can apply the perturbation approach to our quantum measurement problem. We use the joint probability density function introduced in Section~\ref{sec-mainSPI} in its generalized form ${\cal P}_{\zeta} = P(\{ \bm{q}_k \},\{ r_k \}|\qq_0,\zeta)$, a joint PDF of the measurement outcomes and the quantum states given an arbitrary set of constraints $\zeta$. Statistical averages or expectation values using the joint PDF are then given in this form,
\begin{align}\label{eq-pathintA}
\nn\langle {\cal A}\rangle_{\zeta}\equiv &\int \!\!\dd[{\bm q}_k]\dd [r_k] {\cal P}_{\zeta}\, {\cal A},\\
\overset{\delta t \rightarrow 0}{\approx}&{\cal N}\!\!\!\int\!\!\!{\cal D}{\bm q}{\cal D} r {\cal D}{\bm p} \,\exp({\cal S}){\cal  A},
\end{align}
where, in the second line, we have taken the time-continuous limit and written the joint PDF in the path integral form as ${\cal P}_{\zeta} = {\cal N}\!\int\!\! {\cal D}\pp \,\exp({\cal S})$, noting that the time-continuous action is given by Eq.~\eqref{eq-contaction}.

In the following examples, we present the perturbative expansion approach in computing the statistical moments of qubit trajectories. We focus on the case when the qubit Hamiltonian does not commute with the measurement operators ($\Delta \ne 0$) and without a final state constraint.

\subsubsection{Diagrammatic rules for qubit measurement with Rabi oscillation ($\Delta \ne 0$) and with no final state condition}\label{subsec-nonqnd}
The theoretical setup for the qubit with Rabi oscillation is analogous to the one used in Ref.~\cite{Weber2014}, where the qubit Hamiltonian is ${\op H} = (-\Delta/2) \op{\sigma}_x$. Here we can calculate important quantities such as average trajectories and correlation functions, for the case when there is only an initial state fixed and not the final state. For simplicity of the diagrammatic expansions, we make a white noise approximation and variable transformations to the qubit system. 

\begin{table*}
{\renewcommand{\arraystretch}{1.8}
\begin{tabular}{  |>{\centering\arraybackslash} m{2cm} |>{\centering\arraybackslash} m{3.6cm} | >{\centering\arraybackslash} m{3.8cm}|  >{\centering\arraybackslash} m{2.5cm} | }
\hline
Type & Labels of vertices & Full forms & Diagrams  \\ \hline
Type 1 (initial) & $p_{u0}, p_{v0}, p_{w0}$ & $\xp_I\! \int_0^T\!\dd t\, \pxp(t)\delta(t) $ & \begin{tikzpicture}[node distance=0.6cm and 0.8cm]
\coordinate (b2);
\coordinate[right=0.5cm of b2] (bp);
\draw[particle] (b2) -- (bp);
\draw (bp) circle (.06cm);
\end{tikzpicture} \\ \hline
 Type 2 & $\pxp\xp\yp\xi$, $\pxp\xp\zp\xi$, $\pyp\yp\yp\xi$, $\pyp\yp\zp\xi$, $\pzp\zp\yp\xi$, $\pzp\zp\zp\xi$& $\alpha \!\int_0^T \!\dd t \, \pxp(t) \xp(t)\yp(t) \xi(t)$ & \begin{tikzpicture}[node distance=0.3cm and 0.5cm]
\coordinate (b2);
\coordinate[right=0.5cm of b2] (bp);
\coordinate[below=0.5cm of bp](bpp);
\coordinate[right=of bp](c0);
\coordinate[below=0.2cm of c0](c1);
\coordinate[above=0.2cm of c0](c2);
\draw[particle] (b2) -- (bp);
\draw[gluon](bp)--(bpp);
\draw (bp) circle (.06cm);
\draw[particle2](bp) sin (c1);
\draw[particle2](bp) sin (c2);
\end{tikzpicture} \\ \hline
Type 3 & $\pxp \xi$, $\pyp \xi$, $\pzp \xi$ & $\kappa_1\! \int_0^T \!\dd t\, \pxp(t)\xi(t)$ & \begin{tikzpicture}[node distance=0.4cm and 0.5cm]
\coordinate (b2);
\coordinate[right=of b2] (bp);
\coordinate[below=of bp](bpp);
\draw[particle] (b2) -- (bp);
\draw[gluon](bp)--(bpp);
\draw (bp) circle (.06cm);
\end{tikzpicture} \\ \hline 
\end{tabular}}
\caption{We show the 12 terms in the interaction action Eq.~\eqref{eq-interS} (interaction vertices) classified into three types. The second column shows the shorthand labels of the vertices, simply the variables contained in each of interaction terms, while the third column presents examples of their full integration forms. In the fourth column, we illustrate the vertices as connecting edges. The direction of the arrows shown on solid edges is from the $p_{u,v,w}$ variables to the $u,v,w$ variables representing the propagators $G_{u,v,w}$. The wavy lines are for the connection between noise variables. The number of edges (in-coming, out-going, and curly edges) around the vertex (small circle) corresponds to the number of connections required for the vertex.}
\label{tb-vertices}
\end{table*}

We make an idealized white noise limit on the qubit measurement. The variance of the measurement readout distribution $P(r|z)$ is assumed to be very broad, which in this case means $\tau_m \gg \delta t$, justifying an approximation of the two Gaussian distribution in Eq.~\eqref{eq-probr} to a single Gaussian distribution, $P(r | z) \approx (\delta t/2 \pi \tau_m)^{1/2}e^{-(r-z)^2 \delta t/2\tau_m}$ (see Appendix~\ref{app-whitenoise} for more detail). The measurement readout is then approximated to be its mean plus a noise, $r = z + \sqrt{\tau_m}\, \xi$, where $\xi$ is the Gaussian noise with variance $\delta t^{-1}$, independent of the qubit state $z$. Because the nature of this noise is highly fluctuating, in the derivation of the state update equations, we need to keep an expansion up to a second order in $\delta t$, replacing $r^2 \delta t^2 \sim \tau_m \delta t$ \cite{BookGardiner}. This leads to the It\^{o} stochastic differential equations \cite{Korotkov1999,Korotkov2001},
\begin{subequations}\label{eq-itoform}
\begin{align}
{\dot x} &= - \Gamma\, x - x\,z\,\xi/\tau_m^{1/2},\\
{\dot y} &= - \Gamma\, y + \Delta z - y\,z\,\xi/\tau_m^{1/2},\\
{\dot z} &=  - \Delta\, y + (1-z^2)\,\xi/\tau_m^{1/2},
\end{align}
\end{subequations}
where $x,y,z$ are Bloch sphere coordinates for the qubit and a dephasing rate $\Gamma$ is now a total dephasing rate $\Gamma = \gamma + 1/2 \tau_m $. The white noise $\xi$ has a Gaussian probability distribution $P(\xi) = (\delta t/2 \pi)^{1/2}\exp(-\xi^2 \delta t/2)$.

Before substituting the state update above into the action of the path integral, we can make changes in the variables of the system in order to simplify the later perturbation process. This is to avoid infinite series related to the linear terms in Eqs.~\eqref{eq-itoform}. We define a new set of variables $\xp, \yp, \zp$ where $ \{ \xp, \yp, \zp \} = {\bm Q}^{-1} \cdot \{ x,y,z\}$ and ${\bm Q}$ is a matrix that diagonalizes the linear terms of Eqs.~\eqref{eq-itoform},
\begin{equation}\label{eq-transformQ}
{\bm Q}^{-1}\cdot
\left(
\begin{matrix} -\Gamma & 0 & 0 \\ 0  & - \Gamma & +\Delta \\ 0 & - \Delta & 0 \end{matrix} 
\right) \cdot {\bm Q} = 
 \left( \begin{matrix} \diel_1& 0 & 0 \\ 0 & \diel_2 & 0 \\ 0 & 0 & \diel_3 \end{matrix} \right).
\end{equation}
The eigenvalues are $\diel_1 = -\Gamma$, $\diel_2 = -(\Gamma+\Omega)/2$ and $\diel_3 = -(\Gamma-\Omega)/2$, where we define $\Omega = \sqrt{\Gamma^2 - 4 \Delta^2}$. The diagonalizing matrix ${\bm Q}$ and the transformation of the system variables $x,y,z$ are described in the matrix equation,
\begin{equation}\label{eq-transformQ2}
\left( \begin{matrix} x \\ y \\ z \end{matrix} \right) = 
\left(
\begin{matrix} 1 & 0 & 0 \\ 0  & \frac{\Gamma + \Omega}{2 \Delta}& \frac{\Gamma-\Omega}{2 \Delta} \\ 0 & 1 & 1 \end{matrix} 
\right) \cdot \left( \begin{matrix} u \\ v \\ w \end{matrix} \right) ,
\end{equation}
which gives the following transformation: $x = \xp$, $y=\yp\,(\Gamma+\Omega)/2 \Delta +\zp\,(\Gamma-\Omega)/2 \Delta$, and $z = \yp+\zp$, and the inverse transformation: $\xp = x$, $\yp = (2 y \Delta -\Gamma z+\Omega z)/(2 \Omega)$, and $\zp = (-2 y \Delta + \Gamma z+ \Omega z)/(2 \Omega)$.

The stochastic differential equations \eqref{eq-itoform} with this new set of variables are,
\begin{subequations}\label{eq-transformSDE}
\begin{align}
\dot{\xp}& = \diel_1\, \xp +\alpha\, \xp(\yp+\zp) \, \xi + \kappa_1\,\xi , \\ 
\dot{\yp}&= \diel_2 \,\yp + \alpha\, \yp(\yp+\zp) \, \xi + \kappa_2\, \xi , \\ 
\dot{\zp}&= \diel_3 \,\zp + \alpha\, \zp (\yp +\zp)  \,\xi +\kappa_3 \,\xi,
\end{align}
\end{subequations}
where we have defined $\alpha = -\,1/\tau_m^{1/2}$ and $\kappa_1=0$, $\kappa_2=(-\Gamma+\Omega)/(2 \tau_m^{1/2}\Omega)$, $\kappa_3 =(\Gamma+\Omega)/(2 \tau_m^{1/2}\Omega) $.

Now we are ready to construct the action of the path integral, substituting the update equations \eqref{eq-transformSDE} and the log-likelihood function ${\cal F}[\qq,\xi] = - \frac{1}{2}\xi^2 $ into the action Eq.~\eqref{eq-contaction}. We write the action in two separated terms, ${\cal S} = {\cal S}_F + {\cal S}_I$, where
\begin{subequations}\label{eq-actionS}
\begin{align}
\nn{\cal S}_F =\int_0^T\!\!\! \dd t \big\{ \! - & \pxp \,(\dot{\xp}- \diel_1 \xp) -  \pyp \,(\dot{\yp}-\diel_2 \yp) \\
\label{eq-freeS}&-  \pzp \, (\dot{\zp}-\diel_3 \zp) -\xi^2/2 \big\},\\
\nn {\cal S}_I = \int_0^T \!\!\!\dd t \big\{B &
+  \alpha\, \pxp \,\xp(\yp+\zp) \, \xi + \kappa_1\,\pxp\,\xi  \\
\nn&+ \alpha\, \pyp\,\yp(\yp+\zp) \, \xi + \kappa_2\,\pyp\, \xi \\
\label{eq-interS}&+ \alpha\, \pzp\,\zp (\yp +\zp)  \,\xi +\kappa_3 \,\pzp\,\xi \big\},
\end{align}
\end{subequations}
are the free and interaction actions, respectively. We note that the additional term $B$ is not merely a time-continuous form of the first term in Eq.~\eqref{eq-actiongenS-edit} (we only consider the initial condition). This is because the actual initial condition term $-\pp_{-1}\cdot (\qq_0-\qq_I)$ can be removed simply by integrating over the initial variable $\qq_0$, forcing the condition $\{ x(0), y(0), z(0) \} = \{ x_I, y_I, z_I \}$ to the variables at time $t_0 = 0$. In place of the initial term, when we write the free action Eq.~\eqref{eq-freeS} in this form, ${\cal S}_F = - \int \!\!\dd t \dd t'\bxt(t) G^{-1}(t,t') \bxx(t')$, there will be leftover terms which contribute to the term $B$, resulting in,
\begin{align}
\nn B =&  \int \!\!\dd t\,\xp_I\pxp(t)\delta(t) + \int \!\!\dd t\,\yp_I\pyp(t)\delta(t) \\
&+\int \!\!\dd t\, \zp_I \pzp(t) \delta(t).
\end{align}
This can be easily shown with a discretized version of the path integral, which is presented in more detail in Appendix~\ref{app-sourceterm}. We note also that, in the rest of the paper, the time integral is always from $t=0$ to $t=T$, unless stated otherwise.

The propagators (or Green's functions) are computed from the inverse Green's functions, which in this particular case are $G^{-1}_k(t,t') =\delta(t-t') \left(\frac{\dd}{\dd t'} - \diel_{k}\right)$, where $k =1,2,3$ (or $\xp,\yp,\zp$) for the first three terms in Eq.~\eqref{eq-freeS}, and $G_{\xi}^{-1} = \delta(t-t')$  for the noise term. With an identity relation, $\int \dd t'' G^{-1}(t, t'')G(t'',t') = \delta (t-t')$, the propagators (Green's functions, or two-point correlation functions) are given by,
\begin{subequations}\label{eq-propagators}
\begin{align}
\langle \xp(t) \pxp(t') \rangle_F  &= G_{\xp}(t,t') = \Theta(t- t')e^{\diel_1(t-t')},\\
\langle \yp(t) \pyp(t') \rangle_F  &= G_{\yp}(t,t') = \Theta(t- t')e^{\diel_2(t-t')},\\
\langle \zp(t) \pzp(t') \rangle_F  &= G_{\zp}(t,t') = \Theta(t- t')e^{\diel_3(t-t')},\\
\langle \xi(t) \xi(t') \rangle_F  &= G_{\xi}(t,t') = \delta(t- t'),
\end{align}
\end{subequations}
where $t>t'$ for the first three lines. It is important to note that $\Theta(t)$ is a left continuous Heaviside step function, i.e., $\Theta(0)=0$ and $\lim_{t\rightarrow 0^{+}}\Theta(t) = 1$, causing the two-point correlation functions for $\xp,\yp,\zp$ to vanish when $t \le t'$. This is a result of the It\^{o} interpretation we chose in writing the state update equation in Eq.~\eqref{eq-transformSDE}. This can also be verified with the discretized version of the path integral and is presented in Appendix~\ref{app-heaviside}.

To compute quantities such as an expectation value $\langle {\cal A} \rangle = \langle {\cal A}\, e^{{\cal S}_I} \rangle_F$ where ${\cal A}$ is an arbitrary function written in the form of Eq.~\eqref{eq-expA}, one needs to expand the exponential $e^{{\cal S}_I}$ in a power series of ${\cal S}_I$, and looks for terms that contribute to its result. Fortunately, from the two-point correlation functions Eq.~\eqref{eq-propagators}, we know that system variables $\xp,\yp,\zp$ can only connect with the their conjugate variables at earlier times, e.g., $\xp(t_1)$ can only connect with $\pxp(t_2)$ if $t_1 > t_2$. Knowing this helps predict which terms will contribute to the sum of the expansion. For example, to show that $\langle 1 \rangle = \langle e^{{\cal S}_I} \rangle_F = 1$, we look at terms in the interaction action Eq.~\eqref{eq-interS} and see that all of them contain exactly one conjugate variable in each. Looking at the propagators in Eqs.~\eqref{eq-propagators}, we know that it is impossible for any higher order terms in the expansion of $e^{{\cal S}_I}$ to have all conjugate variables matched with system variables at their later times. This is because every time we want to find terms with system variables at later times, there will be at least one more unmatched conjugate variable appearing. Therefore, there are no other contributions apart from 1, resulting in $\langle e^{{\cal S}_I} \rangle_F = 1$.

In order to make the calculation of moments $\langle {\cal A} \rangle =\langle {\cal A} e^{{\cal S}_I}\rangle_F$  more systematic, we develop diagrammatic rules to ease the process of keeping track of the perturbative expansion. Each term in the expansion of $e^{{\cal S}_I}$ is a combination of 12 additive terms shown in the interaction action Eq.~\eqref{eq-interS}, with repeated appearance also possible. The 12 interaction terms (\textit{interaction vertices}) are shown in Table~\ref{tb-vertices} with their labeling names and examples of their full integral forms, where we use the latter to characterize the vertices into three types shown as the three rows. At the end, all combinations of the interaction vertices (all additive terms in the expansion of $e^{{\cal S}_I}$) are then multiplied with the function of interest ${\cal A}$ before taking the free expectation $\langle \cdots \rangle_F$. The functional ${\cal A}$ is a function of system variables and noise variables (i.e., $\xp,\yp, \zp, \xi$) at all times. These variables are called \textit{ending} vertices. Let us use the word \textit{combination} for an individual term combining interaction vertices (from additive terms in the expansion $e^{{\cal S}_I}$) and the ending vertices (from the function ${\cal A}$). One can determine which combinations are non-vanishing and compute their values by following these diagrammatic rules:
\begin{itemize}
\item A non-vanishing combination of vertices should contain equal number of each of the system variables ($\xp,\yp,\zp$) and their conjugates ($\pxp, \pyp, \pzp$), and there should be even number of the noise variables ($\xi$).
\item The system variable can only be connected to its conjugate at earlier time, while the noise variable $\xi$ connects to another noise variable at the same time. The connections are presented as `edges', or lines joining vertices. The number of edges for each vertex is equal to the number of variables it contains (graphical representations are shown in Table~\ref{tb-vertices}, in the column of `Diagrams'). A non-vanishing combination can have any number of vertices, but edges have to all be connected.
\item A non-vanishing combination is presented by connected diagrams with the ending vertices (variables with their time arguments being in the range of $t \in (0, T]$) on the most left and their time arguments clearly stated. The time ordering decreases from left to right, with variables at the same time lining up in the same vertical line, and the vertices with initial conditions ($t=0$) on the most right.
\item A result of the free expectation value ($\la \cdots \ra_F$) for each combination is computed by taking into account the constants (i.e., $\xp_I, \yp_I, ..., \alpha, \kappa_1, ...$) and integrals attached to each of the interaction vertices, the propagators (the Green's function), and the numbers of possible ways to connect diagrams. From the point of view of graphical diagrams, edges represent propagators, and points common to more than one edge correspond to integrals over time. Moreover, if there are $m$ repeated vertices in a combination, it needs to be multiplied by a factor $\frac{1}{m!}$ (resulting from coefficients in the expansion of $e^{{\cal S}_I}$). However, these factors usually cancel with the number of possible ways to connect the repeated vertices.
\end{itemize}

\begin{figure}
\includegraphics[width=5.8cm]{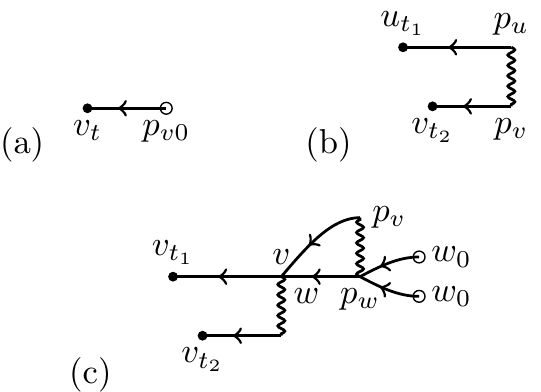}
\caption{Examples of diagrams, where the labels indicating the variables contained in each of the vertices (we label only the ones that are necessary). The filled dots represent the ending vertices, while the open dots represent interaction (initial, $t=0$) vertices. (a) An ending vertex connects with an initial vertex ($p_{v0}$). (b) Two ending vertices connect with two interaction vertices ($p_u \xi$ and $p_v \xi$). (c) Two ending vertices connect with six interaction vertices (from left-right, top-bottom order: $p_v v w \xi$, $p_v \xi$, $p_v \xi$, $p_w w w \xi$, $p_{w0}$ and $p_{w0}$). The unequal length of horizontal edges following the ending vertices $v_{t_1}$ and $v_{t_2}$ indicate that $t_1 \ge t_2$.}
\label{fig-exdiagram}
\end{figure}

In Figure~\ref{fig-exdiagram}, we show examples of non-vanishing diagrams showing connections of one and two ending vertices. For each of the diagrams, we can determine its contribution by explicitly writing out its full form and computing its integrals. In the following subsections, we present some of the calculations using these diagrams and rules to estimate statistical averages of the qubit trajectories.

We note that it is also possible to construct different diagrammatic rules based on different formations of the path integrals. For example, one can perform integrals over the readout noise (e.g., the noise $\xi$), transforming the action in Eq.~\eqref{eq-freeS} and \eqref{eq-interS} to a new action that contains only the system variables $u, v, w$. In this case, the diagrams will be completely different from the one we presented above, but still giving exactly the same results for the statistical quantities related to the system variables. Here, we choose to present the diagrammatic rules with explicit noise variables because of the ability to compute correlation functions between the noise and the system states, as shown in section~\ref{sec-noisesyscorr}.

\subsubsection{Small noise approximation and loop expansion}\label{sec-smallnoiseapprox}
The diagrammatic rules help in determining non-vanishing terms, but we need a systematic approach to keep track of the order of expansion, especially when there are infinitely many ways to construct diagrams. We consider the small noise approximation around the saddle point solution mentioned in section~\ref{sec-briefintro}. To obtain the action of the form in Eq.~\eqref{eq-actionsemiclass}, we introduce a small parameter denoted by $\nu$ to the noise term of Eqs.~\eqref{eq-actionS}, controlling the noise variance in the qubit dynamics, leading to a modified action,
\begin{align}\label{eq-semiclassaction}
\nn{\cal S}' =\int_0^T\!\!\! \dd t \big\{ \! - & \pxp \,(\dot{\xp}- \diel_1 \xp) -  \pyp \,(\dot{\yp}-\diel_2 \yp) \\
&-  \pzp \, (\dot{\zp}-\diel_3 \zp) -\xi^2/2 \nu \big\} + {\cal S}_I,
\end{align}
leaving all terms in the interaction action intact. We then rescale the conjugate variables by replacing $p_u \rightarrow p_u/\nu$, $p_v \rightarrow p_v/\nu$ and $p_w \rightarrow p_w/\nu$ (as well as changing variables in the source term, e.g., $J_{u, v,w} \rightarrow J_{u,v,w}/\nu$), noting that the rescaling of the auxiliary variables does not affect the qubit dynamics. Since all terms in the action Eq.~\eqref{eq-semiclassaction}, except for the noise term, have exactly one conjugate variable (see Eqs.~\eqref{eq-actionS}), the resulting action is then in the desired form,
\begin{align}\label{eq-semiclassaction2}
\nn{\cal S}' =\frac{1}{\nu}\bigg[\int_0^T\!\!\! \dd t \big\{ \! - & \pxp \,(\dot{\xp}- \diel_1 \xp) -  \pyp \,(\dot{\yp}-\diel_2 \yp) \\
&-  \pzp \, (\dot{\zp}-\diel_3 \zp) -\xi^2/2 \big\} + {\cal S}_I \bigg],
\end{align}
where $p_{u,v,w}$ are the rescaled conjugate variables.

Following the discussion in section~\ref{sec-briefintro} and the diagrammatic rules presented in the previous section, the order of expansion can be determined by counting the number of edges and vertices of a diagram. Each vertex in a diagram contributes a factor of $1/\nu$ to its result, while each edge (which represents a propagator) contributes a factor of $\nu$. Therefore, each term (diagram) in the expansion carries a factor of $\nu^{E+I-V}$, where $E$, $I$, and $V$ are numbers of its external edges (edges that connect with ending vertices), internal edges, and vertices, respectively. This corresponds to the loop expansion where one instead counts the number of loops from $L = I-V+1$, which leads to another way of writing the order of the expansion $\nu^{E+L-1}$, noting that the number of external edges $E$ is fixed for each set of ending vertices.

We show how to count the power of $\nu$ of the diagrams shown in Figure~\ref{fig-exdiagram}. The numbers of edges, vertices, and loops are as follows: (a) $E = 1$, $I=0$, $V =1$, $L=0$, (b) $E=2$, $I=1$, $V=2$, $L=0$, (c) $E=2$, $I=6$, $V=6$, $L=1$, giving the zeroth(zeroth), zeroth(first), and first(second) orders of loop($\nu$), respectively.
 
\subsubsection{Example A: Average quantum trajectory}\label{sec-averagesol}
As an example of how to use the diagrammatic rules to compute statistical expectation values, we show a derivation of $\langle z(t) \rangle$, an average of the $z$-coordinate of the quantum trajectories. From the variable transformation in Eq.~\eqref{eq-transformQ2}, we know that $z(t) = \yp(t) + \zp(t)$, therefore we can write the expectation value in terms of the free moments and then compute the diagrams from the vertices in Table~\ref{tb-vertices},
\begin{align}\label{eq-avez}
\nn \langle  z(t) \rangle = &\langle  \yp(t)e^{{\cal S}_I} \ra_F + \la \zp(t)e^{{\cal S}_I} \rangle_F, \\
\nn =& \,\begin{tikzpicture}[node distance=0.6cm and 0.8cm]
\coordinate[label=below:$v_t$] (b2);
\coordinate[right=of b2,label=below:$p_{v0}$] (bp);
\draw[particle] (b2) -- (bp);
\draw (bp) circle (.06cm);
\fill[black] (b2) circle (.05cm);
\end{tikzpicture}\, + \,\begin{tikzpicture}[node distance=0.6cm and 0.8cm]
\coordinate[label=below:$w_t$] (b2);
\coordinate[right=of b2,label=below:$p_{w0}$] (bp);
\draw[particle] (b2) -- (bp);
\draw (bp) circle (.06cm);
\fill[black] (b2) circle (.05cm);
\end{tikzpicture},\\
\nn=&\left\langle \yp(t)\,\yp_I \!\!\!\int \!\!\!\dd t'\, \pyp(t')\delta (t') \right\rangle_F \\
&\nn + \left\langle \zp(t)\,\zp_I \!\!\!\int \!\!\!\dd t'\, \pzp(t')\delta (t') \right\rangle_F,\\
\nn=& \,\yp_I \!\!\! \int \!\!\dd t' G_{\yp}(t ,t')\delta(t')+\zp_I \!\!\!\int\!\! \dd t' G_{\zp}(t ,t')\delta(t'), \\
=&\, \yp_I e^{\diel_2 t} + \zp_I e^{\diel_3 t}, \quad \text{where  $t \ge 0$},
\end{align}
where the Green's functions are from Eqs.~\eqref{eq-propagators}. From the types of vertices in Table~\ref{tb-vertices}, there is only one possibility connecting the ending vertices $\yp(t)$ and $\zp(t)$ with $p_{v0}$ and $p_{w0}$, respectively. This is because choosing other vertices will lead to at least one unmatched noise variable $\xi$, which, when trying to pair it with another vertex containing $\xi$, will result in more unmatched conjugate variables.

As a result, the expectation value Eq.~\eqref{eq-avez} is explicitly found by transforming back to the system variables $x,y,z$,
\begin{align}
\langle z(t) \rangle =  e^{-\Gamma t/2}\left( z_I \cosh \frac{\Omega t}{2} + \frac{z_I \Gamma - 2 \Delta y_I}{\Omega}\sinh\frac{\Omega t}{2} \right),
\end{align}
which is exactly the same as the solution we would get from averaging over all possible noise realizations in the It\^{o} stochastic master equations in Eqs.~\eqref{eq-itoform} and solving for an average of $z(t)$. The averages of $x$ and $y$ can be computed in the similar way. We note that the exact solution for the average trajectory involves only the diagrams with zeroth order of the small parameter $\nu$ ($E+I-V = 0$).

Particularly, one can show that the average trajectory coincides with a saddle point solution of the action Eq.~\eqref{eq-semiclassaction} when $\nu \rightarrow 0$. The vanishing functional derivatives of the action over the variables $p_{u,v,w}$ gives the exact same equations Eqs.~\eqref{eq-transformSDE} (though, the noise variable $\xi$ is no longer a stochastic function), while the vanishing functional derivatives over $u,v,w$ leads to the differential equations for conjugate variables,
\begin{subequations}\label{eq-saddlepointp}
\begin{align}
\dot{p}_u & = - \lambda_1 p_u - \alpha (v+w) p_u \xi, \\
\dot{p}_v & = - \lambda_2 p_v - \alpha \{u p_u +  (2 v + w)p_v + w p_w\} \xi, \\
\dot{p}_w & = - \lambda_3 p_w - \alpha \{ u p_u + v p_v + (v + 2 w) p_w \} \xi,
\end{align}
\end{subequations}
where the noise variable is now a solution of extremizing the action over $\xi$ giving,
\begin{align}
\nn \xi = \,\,\nu \,\big\{ &\alpha (v+w)(u p_u + v p_v + w p_w) \\
&+ \kappa_1 p_u + \kappa_2 p_v + \kappa_3 p_w \big\}.
\end{align}
As the noise parameter $\nu$ decreases to zero, as does the noise term $\xi$, the ordinary differential equations of $u,v,w$ are uncoupled from Eqs.~\eqref{eq-saddlepointp} and their solution (a saddle point solution) is then the same as the average trajectory, the solution of Eqs.~\eqref{eq-transformSDE} when $\xi = 0$.


\subsubsection{Example B: Correlation function between system variables and noise variables}\label{sec-noisesyscorr}
Another example is a correlation function between the system variable $z$ and the noise variable $\xi$, given by $\langle z(t_1) \xi(t_2) \rangle=\langle \yp(t_1)\xi(t_2)\rangle+\langle\zp(t_1)\xi(t_2)\rangle$. Since there are now two ending vertices, the situation is more complicated and there are infinitely many ways to connect the diagrams. However, we can compute terms to the lowest order of the loop expansion, which in this case is the zeroth loop (tree-level diagrams). There are in total six contributions to the correlation function, as shown in the following,
\begin{align}\label{eq-noisesyscorr}
\nn \langle z(t_1)& \xi(t_2) \rangle^{(0)}= \langle \yp(t_1)\xi(t_2) e^{{\cal S}_I}\rangle_F^{(0)}+ \la \zp(t_1) \xi(t_2) e^{{\cal S}_I}\ra_F^{(0)} , \\ 
\nn =&
\,\, \begin{tikzpicture}[node distance=0.6cm and 0.8cm]
\coordinate[] (b2);
\coordinate[right=of b2,label=left:$\xi_{t_2}$] (bp);
\coordinate[above=of b2,label=above:$v_{t_1}$](a1);
\coordinate[above=of bp,label=above:$p_{v}$](ap);
\draw[particle] (a1) -- (ap);
\draw[gluon] (ap) --  (bp);
\fill[black] (a1) circle (.05cm);
\fill[black] (bp) circle (.05cm);
\end{tikzpicture}\,+ \!\!\!\!\begin{tikzpicture}[node distance=0.6cm and 0.8cm]
\coordinate[] (b2);
\coordinate[right=of b2,label=left:$\xi_{t_2}$] (bp);
\coordinate[above=of bp,label=below right:$v$,label=above:$v$] (ap);
\coordinate[left=of ap,label=above:$v_{t_1}$] (a1);
\coordinate[right=of ap](cc);
\coordinate[above=0.3cm of cc,label=right:$v_0$](cp);
\coordinate[below=0.1cm of cc,label=right:$v_0$](dp);
\draw[particle] (a1) -- (ap);
\draw[gluon] (ap) --  (bp);
\draw[particle2] (ap) sin  (dp);
\draw[particle] (ap) sin  (cp);
\fill[black] (a1) circle (.05cm);
\fill[black] (bp) circle (.05cm);
\draw (cp) circle (.06cm);
\draw (dp) circle (.06cm);
\end{tikzpicture}+
\!\!\!\!\begin{tikzpicture}[node distance=0.6cm and 0.8cm]
\coordinate[] (b2);
\coordinate[right=of b2,label=left:$\xi_{t_2}$] (bp);
\coordinate[above=of bp,label=below right:$w$,label=above:$v$] (ap);
\coordinate[left=of ap,label=above:$v_{t_1}$] (a1);
\coordinate[right=of ap](cc);
\coordinate[above=0.3cm of cc,label=right:$v_0$](cp);
\coordinate[below=0.1cm of cc,label=right:$w_0$](dp);
\draw[particle] (a1) -- (ap);
\draw[gluon] (ap) --  (bp);
\draw[particle2] (ap) sin  (dp);
\draw[particle] (ap) sin  (cp);
\fill[black] (a1) circle (.05cm);
\fill[black] (bp) circle (.05cm);
\draw (cp) circle (.06cm);
\draw (dp) circle (.06cm);
\end{tikzpicture}\\
\nn&+\!\!\!\begin{tikzpicture}[node distance=0.6cm and 0.8cm]
\coordinate[] (b2);
\coordinate[right=of b2,label=left:$\xi_{t_2}$] (bp);
\coordinate[above=of b2,label=above:$w_{t_1}$](a1);
\coordinate[above=of bp,label=above:$p_{w}$](ap);
\draw[particle] (a1) -- (ap);
\draw[gluon] (ap) --  (bp);
\fill[black] (a1) circle (.05cm);
\fill[black] (bp) circle (.05cm);
\end{tikzpicture}\,+\!\!\!\! \begin{tikzpicture}[node distance=0.6cm and 0.8cm]
\coordinate[] (b2);
\coordinate[right=of b2,label=left:$\xi_{t_2}$] (bp);
\coordinate[above=of bp,label=below right:$w$,label=above:$w$] (ap);
\coordinate[left=of ap,label=above:$w_{t_1}$] (a1);
\coordinate[right=of ap](cc);
\coordinate[above=0.3cm of cc,label=right:$w_0$](cp);
\coordinate[below=0.1cm of cc,label=right:$w_0$](dp);
\draw[particle] (a1) -- (ap);
\draw[gluon] (ap) --  (bp);
\draw[particle2] (ap) sin  (dp);
\draw[particle] (ap) sin  (cp);
\fill[black] (a1) circle (.05cm);
\fill[black] (bp) circle (.05cm);
\draw (cp) circle (.06cm);
\draw (dp) circle (.06cm);
\end{tikzpicture}+
\!\!\!\!\begin{tikzpicture}[node distance=0.6cm and 0.8cm]
\coordinate[] (b2);
\coordinate[right=of b2,label=left:$\xi_{t_2}$] (bp);
\coordinate[above=of bp,label=below right:$v$,label=above:$w$] (ap);
\coordinate[left=of ap,label=above:$w_{t_1}$] (a1);
\coordinate[right=of ap](cc);
\coordinate[above=0.3cm of cc,label=right:$w_0$](cp);
\coordinate[below=0.1cm of cc,label=right:$v_0$](dp);
\draw[particle] (a1) -- (ap);
\draw[gluon] (ap) --  (bp);
\draw[particle2] (ap) sin  (dp);
\draw[particle] (ap) sin  (cp);
\fill[black] (a1) circle (.05cm);
\fill[black] (bp) circle (.05cm);
\draw (cp) circle (.06cm);
\draw (dp) circle (.06cm);
\end{tikzpicture},\\
\nn=& \,\kappa_2 \!\!\int \!\!\dd t' G_{\yp}(t_1,t') G_{\xi}(t',t_2)\\
\nn&+\alpha \yp_I^2\!\! \int\!\! \dd t' G_{\yp}(t_1, t')G_{\xi}(t',t_2)G_{\yp}(t', 0)^2 + \cdots,\\
\nn=&\, e^{\diel_2 t_1}\left(\kappa_2 e^{-\diel_2 t_2}+\alpha \yp_I^2 e^{\diel_2 t_2}+\alpha \yp_I \zp_I e^{\diel_3 t_2}\right)\\
&+e^{\diel_3 t_1}\left(\kappa_3 e^{-\diel_3 t_2}+\alpha \zp_I^2 e^{\diel_3 t_2}+\alpha \yp_I \zp_I e^{ \diel_2 t_2}\right),
\end{align}
where the superscript $(0)$ in the first line indicates the number of loops in the expansion, and, in the third line, we show only the integral form of the first two diagrams. This result involves diagrams with first order of the small noise parameter $\nu$. The definitions of the parameters such as $u_I, \kappa_2, \diel_2$ are defined in the discussion of Eqs.~\eqref{eq-transformQ}-\eqref{eq-transformQ2}.

We note that this result Eq.~\eqref{eq-noisesyscorr} is valid for $t_1 > t_2$ and it is zero otherwise, because of the special properties of the left continuous Heaviside step function $\Theta(t)$. This can be interpreted as the noise being correlated with the qubit state at later times but not with the state at earlier times. This property of the system variable and noise the correlation function is also mentioned in Ref.~\cite{Korotkov2001-2}.

\subsubsection{Example C: Correlation function between two system variables}
Let us next consider correlation functions between system variables at two points in time, such as $\la z(t_1) z(t_2) \ra$ and $\la y(t_1) z(t_2) \ra$. Both of them can be written in terms of correlation functions between the transformed system variables $v,w$, for example, the $z$-$z$-correlation is, $\langle z(t_1) z(t_2) \rangle=\langle \yp(t_1)\yp(t_2)\rangle+\langle\yp(t_1)\zp(t_2)\rangle+\langle\zp(t_1)\yp(t_2)\rangle+\langle\zp(t_1)\zp(t_2)\rangle$. 
As in the previous example, we keep terms up to the lowest order of loops, which is the zeroth order.

Each term of the correlation functions, for example $\langle \yp(t_1)\yp(t_2)\rangle$, involves 10 different diagrams, thus there are in total $4 \times 10$ diagrams for computing the $z$-$z$ correlation function. We present in the following three samples of diagrams and their integral forms for the correlation function $\la v(t_1)v(t_2) \ra$,
\begin{align}\label{eq-corrfourth}
\nn \langle \yp(t_1)& \yp(t_2) \rangle^{(0)} = \langle \yp(t_1)\yp(t_2)e^{{\cal S}_I}\rangle_F^{(0)},  \\ 
\nn =&
\,\, \begin{tikzpicture}[node distance=0.6cm and 0.8cm]
\coordinate[label=below:$v_{t_2}$] (b2);
\coordinate[right=of b2,label=below:$p_{v0}$] (bp);
\coordinate[above=of b2,label=above:$v_{t_1}$](a1);
\coordinate[above=of bp,label=above:$p_{v0}$](ap);
\draw[particle] (b2) -- (bp);
\draw[particle] (a1) -- (ap);
\draw (bp) circle (.06cm);
\draw (ap) circle (.06cm);
\fill[black] (b2) circle (.05cm);
\fill[black] (a1) circle (.05cm);
\end{tikzpicture}\,+\begin{tikzpicture}[node distance=0.6cm and 0.8cm]
\coordinate[label=below:$v_{t_2}$] (b2);
\coordinate[right=of b2,label=below:$p_v$] (bp);
\coordinate[above=of bp,label=above:$p_v$] (ap);
\coordinate[left=1.1cm of ap,label=above:$v_{t_1}$] (a1);
\draw[particle] (a1) -- (ap);
\draw[particle] (b2) -- (bp);
\draw[gluon] (ap) --  (bp);
\fill[black] (a1) circle (.05cm);
\fill[black] (b2) circle (.05cm);
\end{tikzpicture}\,+
\,\,\begin{tikzpicture}[node distance=0.6cm and 0.8cm]
\coordinate[label=below:$v_{t_2}$] (b2);
\coordinate[right=of b2,label=below:$p_v$] (bp);
\coordinate[above=of bp,label=below right:$v$,label=above:$v$] (ap);
\coordinate[left=1.1cm of ap,label=above:$v_{t_1}$] (a1);
\coordinate[right=of ap](cc);
\coordinate[above=0.3cm of cc,label=right:$v_0$](cp);
\coordinate[below=0.1cm of cc,label=right:$v_0$](dp);
\draw[particle] (a1) -- (ap);
\draw[particle] (b2) -- (bp);
\draw[gluon] (ap) --  (bp);
\draw[particle2] (ap) sin  (dp);
\draw[particle] (ap) sin  (cp);
\fill[black] (a1) circle (.05cm);
\fill[black] (b2) circle (.05cm);
\draw (cp) circle (.06cm);
\draw (dp) circle (.06cm);
\end{tikzpicture}\\
\nn & + \text{(7 more similar diagrams)},\\
\nn=& \,\la \yp(t_1) \ra \la \yp(t_2) \ra \\
\nn&+\kappa_2^2\!\! \int\!\! \dd t' \dd t'' G_{\yp}(t_1, t')G_{\yp}(t_2, t'')G_{\xi}(t', t'')\\
\nn&+\alpha \kappa_2 \yp_I^2 \!\!\int\!\! \dd t' \dd t''\big\{G_{\yp}(t_1,t')G_{\yp}(t_2,t'')\\
&\qquad \quad G_{\xi}(t', t'')G_{\yp}(t',0)G_{\yp}(t',0)\big\} + \cdots,
\end{align}
where, apart from the first trivial diagram, each initial vertex ($\yp_{t_1}$ or $\yp_{t_2}$) can connect to three possible interaction vertices ($\pyp \xi$, $\pyp \yp \yp \xi$, or $\pyp \yp \zp \xi$).  As a result, there are total of $9+1$ possible ways, and they are presented in Appendix~\ref{app-fullform} (all diagrams except the first one are of the first order of $\nu$). The similar kind of calculation is applied to the other correlation functions $\langle \yp(t_1)\zp(t_2)\rangle$, $\langle\zp(t_1)\yp(t_2)\rangle$ and $\langle\zp(t_1)\zp(t_2)\rangle$. 

The correlation function between $y$ and $z$ can be calculated in a similar way, with the same types of diagrams, but with extra prefactors $\beta_1$ and $\beta_2$,
\begin{align}
\nn \langle y(t_1)& z(t_2) \rangle = \beta_1 \la v(t_1) v(t_2) \ra + \beta_1 \la v(t_1)w(t_2) \ra\\
\nn&+  \beta_2 \la w(t_1) v(t_2) \ra+ \beta_2 \la w(t_1)w(t_2) \ra,
\end{align}
where the coefficients $\beta_1 = (\Gamma+\Omega)/2 \Delta$ and $\beta_2 = (\Gamma-\Omega)/2 \Delta$ come from the transformation of variables shown in Eq.~\eqref{eq-transformQ2}.

\begin{figure*}
\includegraphics[width=17.6cm]{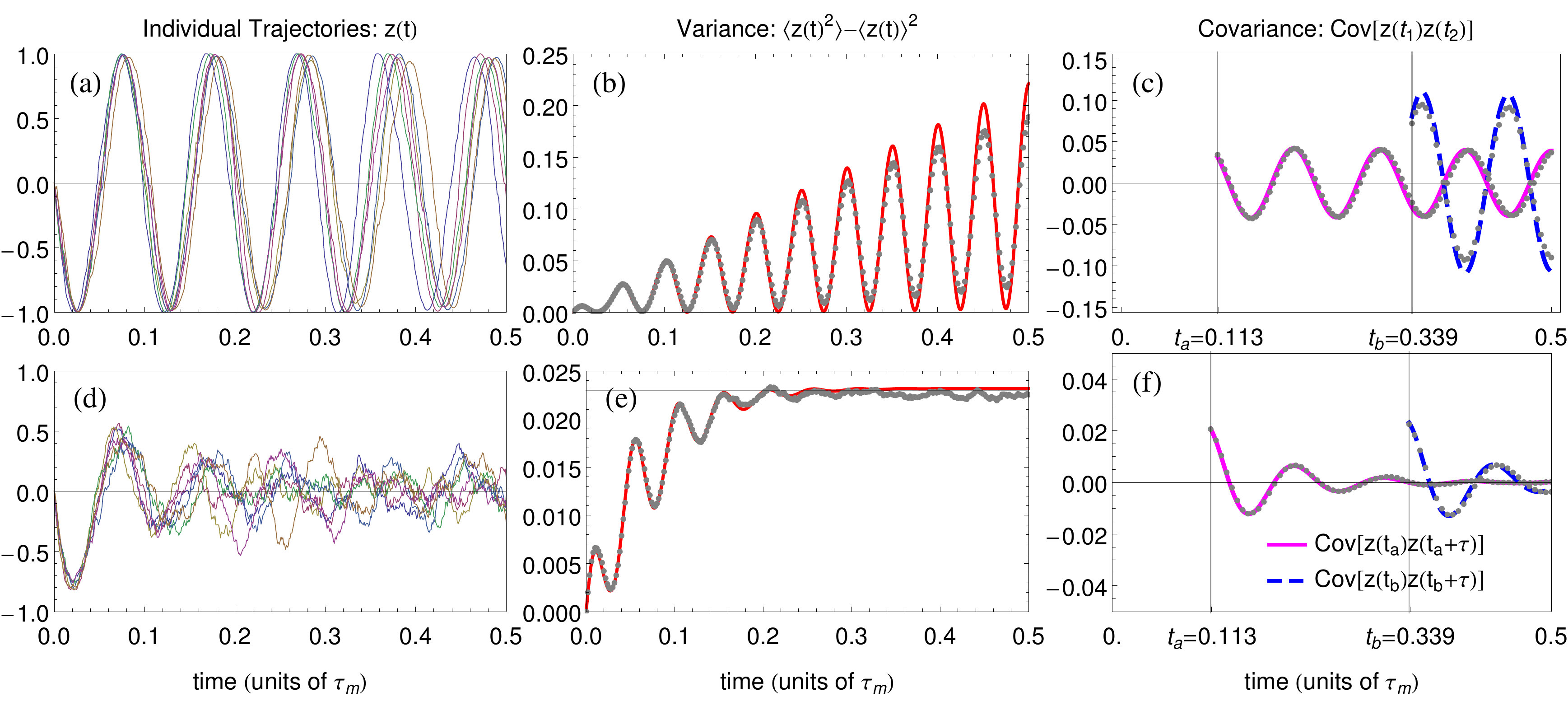}
\caption{(Color online) The variances and covariances for the $z$-coordinate of the qubit trajectories, comparing the theoretical solutions computed up to fourth order of vertices and numerical simulations (based on $ 10^5$ trajectories). The qubit Rabi frequency used in these cases is $\Delta = 20 \pi \, \tau_m^{-1} $, and the total time is $T=0.5\, \tau_m$. The regime presented in the upper row of panels is the case of efficient measurement with no extra decoherence (i.e., $\gamma=0$, $\eta \equiv 1/(2 \tau_m \Gamma) = 1$), while the regime presented in the lower row case is with low efficiency $\eta = 0.02$. The initial state used here is $(x_I, y_I, z_I) = (0, 1,0)$. (a), (d) The simulated individual trajectories for both regimes. (b), (e) The theoretical variances $\la z(t)^2 \ra - \la z(t) \ra^2$ (red curves) along with the numerical data (gray dots). In the panel (e), the variance increases from zero at the initial point $t=0$ and then converges to a value of $(\Gamma^2+\Delta^2)/(2 \Gamma \Delta^2 \tau_m) \approx 0.023 $. (c), (f) The covariances between $z$ at time $t_1 = t_a = 0.113\, \tau_m$ and time $t_2 = t_a+\tau$ (solid magenta curve), and between $z$ at time $t_1 = t_b=0.339\, \tau_m$ and time $t_2 = t_b+\tau$ (dashed blue curve), are presented along with results from the numerical simulation (gray dots).}
\label{fig-corrlimit}
\end{figure*}

\subsubsection{Discussion and comparison with numerical simulation}
We compare the theoretical results, specifically the $z$-$z$ correlation computed in the previous section, with numerically simulated quantum trajectories. Although, the solution of $\la z(t_1) z(t_2) \ra$ is not explicitly presented here in this paper, as it is too lengthy, it can be computed in a similar way as shown in Eq.~\eqref{eq-corrfourth} and also in Appendix~\ref{app-fullform}.

We show in Figure~\ref{fig-corrlimit}(a) and \ref{fig-corrlimit}(d) sampled individual trajectories generated from the Monte Carlo method (Appendix~\ref{app-numer}), in the regime where $T < \tau_m$. The theoretical variances are computed from the $z$-$z$ correlation function, and they are shown in panels (b) and (e) along with the variances from the numerical trajectories. In the top panel, for efficient detection $(\eta \equiv 1/(2 \tau_m \Gamma) = 1)$, the numerical variance increases in time and gets saturated (not shown) at around a value of $0.5$ for a long enough time, indicating that the qubit states at the later times are distributed throughout the perimeter of the $y$-$z$ plane of the Bloch sphere. The variance from the tree-level diagram approximation fails to exactly capture the long-time behaviour, as we can see that the discrepancy in the panel (b) starts to grow as time increases. However, for the inefficient detection case, $\eta = 0.02$ in Figure~\ref{fig-corrlimit}(e), we can see that the theoretical approximation can explain the behaviour quite well, predicting the saturated variance at a value of $(\Gamma^2+\Delta^2)/(2 \Gamma \Delta^2 \tau_m) \approx 0.023$. To compute this quantity theoretically, one takes a limit $t \rightarrow \infty$ of the calculated variance $\la z(t)^2 \ra - \la z(t) \ra^2$.

For the correlation function at two different times, we define a covariance as $\text{Cov}[z(t_1)z(t_2)] \equiv \langle z(t_1)z(t_2) \rangle - \langle z(t_1) \rangle \langle z(t_2) \rangle$, and show in Figure~\ref{fig-corrlimit}(c) and \ref{fig-corrlimit}(f) its numerical and theoretical comparisons. We plot the covariance for $t_1 = t_{a,b}$ and $t_2 = t_{a,b} + \tau$ where $\tau$ has its value ranged from $0$ to $T-t_{a,b}$. For the efficient detection case, shown in the panel (c), the agreement between the theory and the simulation is better for the short time case, $t_a = 0.113\,  \tau_m$. For the inefficient detection case, panel (f), the agreement is excellent for both $t_a$ and $t_b$.

Since our theoretical results are derived with the small noise expansion around the saddle point solution (or the average solution, see section~\ref{sec-averagesol}) using $\nu$ as an expansion tracking parameter, the diagrammatic approximation method works well whenever the qubit trajectories are narrowly distributed around its average. This happens in the short-time regime when the diffusion is still small from the initial state, as well as in the regime when the dynamics is strongly suppressed by the dephasing mechanism (when the detection efficiency is low). We note that for the long-time limit, one needs to compute higher order terms which contain more complex diagrams. However, there are other approaches, such as using the qubit master equation to approximate the correlation functions in the stationary limit and calculate spectral densities of the measurement readouts. These can be found in the works on continuous measurement of mesoscopic electronics such as in Refs.~\cite{KorotkovAverin2001,Korotkov2001-2,Goan2001,Korotkov2003}. Comparing to these approaches, our results give much more accurate correlation functions at short times.

\section{Stochastic path integral for continuous measurement with feedback}\label{sec-feedback}
We can extend the discussion of the stochastic path integral formalism to its application to the system under continuous measurement with feedback control. The feedback loop consists of getting information about the system state via the measurement, and feeding back a control signal to the system in order to alter the state as desired. One of the most intuitive models of the feedback loop is that the system Hamiltonian changes as a function of the system state, which then can be written as a function of the most recent measurement readout. For example, in our qubit measurement case, taking into account additional time delay $\tau_d$, the Hamiltonian at any time $t$ can be written as a function of the measurement readout in the past, $H_{fb}(r_{t-\tau_d}, t) = \epsilon(r_{t-\tau_d})\op{\sigma}_z/2 - \Delta(r_{t-\tau_d})\op{\sigma}_x/2$, where the parameters $\epsilon$ and $\Delta$, as defined in Section~\ref{sec-qubit}, are now functions of the measurement readout $r_{t-\tau_d}$ at time $t-\tau_d$.

We consider an ideal case with instantaneous feedback $\tau_d \approx 0$, i.e., the measurement readout at time $t$ immediately changes the system parameters which are used in computing the state update at the time. This way, the formulation of the stochastic path integral as time-local, its action's extremization equations, and the diagrammatic expansion we presented so far are perfectly applicable. The only modification needed is to treat the readout-dependent Hamiltonian parameters as functions of $r_t = r(t)$ such as $\epsilon(r_t)$ and $\Delta(r_t)$ (or as functions of the quantum state at the time). In this section, we present a few examples of the continuous measurement of a qubit with feedback, one with a linear feedback in the form $\Delta(r_t) = \Delta_0 + \Delta_1 r_t $, and another with a state-dependent linear feedback to stabilize a qubit's oscillation.

\subsection{Linear feedback, most likely path, and its phase space diagram}
Let us consider an instantaneous feedback loop in a qubit measurement introduced in section~\ref{sec-qubit}, with $\epsilon = 0$ and the qubit Rabi frequency being a function of the  measurement readout. We assume a linear form of the Rabi frequency as
\begin{align}\label{eq-feedfreq}
\Delta_{fb}(r_t) = \Delta_0 +\Delta_1 r_t,
\end{align}
where $r_t = r(t)$ is the readout as a function of time and $\Delta_0,\Delta_1$ are constant parameters characterizing the bare Rabi oscillation and the linear feedback, respectively. We will see later in this subsection that this type of feedback can stabilize pre-determined (arbitrary) quantum states. 

This feedback loop in Eq.~\eqref{eq-feedfreq} has the advantage of fast processing because the qubit Hamiltonian depends directly on the value of the measurement readout, requiring no knowledge of the qubit state. For simplicity, we limit our discussion to an ideal case in which the qubit is measured with an efficient detector and no extra environment dephasing. In this regime, we can re-parametrize the qubit Bloch vector into a single state parameter $\theta$ where $z = \cos\theta$ and $y = \sin \theta$ (we set $x =0$ in this case). We note that, with this linear feedback Eq.~\eqref{eq-feedfreq}, a typical step size for the qubit state defined as $\Delta \theta$ has average drift $\la \Delta \theta \ra \sim \delta t$ and diffusion $\la (\Delta \theta)^2 \ra \sim \delta t$ of the first order of the time step. Thus, the step size is still small.

The statistics of quantum trajectories with this instantaneous feedback can be investigated following the same procedure presented in previous sections for the no-feedback qubit measurement. We start with writing the joint probability distribution of the trajectories, and then transform it into a path integral form. The action of the stochastic path integral with the feedback Rabi frequency $\Delta_{fb}(r_t)$ is given by,
\begin{align}\label{eq-fbaction}
\nn{\cal S}_{fb} = \int_0^T\!\!\! \dd t \big\{ & -p_{\theta}(\dot{\theta} - \Delta_0 -\Delta_1 \, r + r \sin\theta \,/\tau_m) \\
&- (r^2 - 2 \,r \cos \theta +1)/2 \tau_m \big\},
\end{align}
where $p_{\theta}$, $\theta$, and $r$ are functions of time, omitting the time argument. This is a generalization of the action given in Ref.~\cite{Chantasri2013} where the quantum jump in qubit measurement is analyzed. We note here that in Eq.~\eqref{eq-fbaction}, we have chosen to use the $\delta t$-expansion in both the state update and the readout probability distribution, because we are interested in investigating the action-extremized solutions, the optimal paths of the system.

\begin{figure*}[t]
\includegraphics[width=17.9cm]{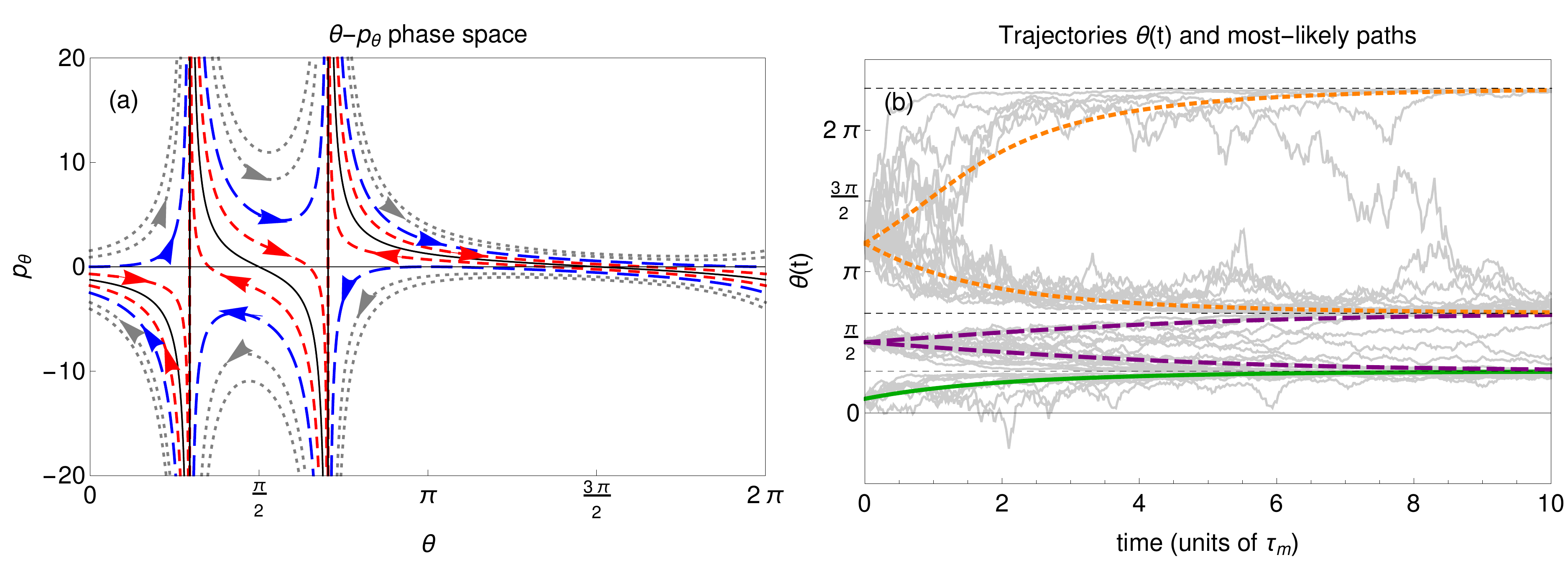}
\caption{(Color online) The phase space portrait and the most likely paths for the continuous measurement with linear feedback. (a) The $\theta$-$p_{\theta}$ phase space portrait for the pure linear feedback case ($\Delta_0  = 0$) with $\Delta_1 = 0.8\, \tau_m^{-1}$, where the collapse angles shown as two vertical lines are at $\theta_{s1,s2} \approx 0.3 \pi, 0.7 \pi$. The curves represent the most likely paths for different stochastic energies: $E= 0$ (long dashed blue), $E = E_c = -1/2\tau_m = -0.5$ (solid black), $ E_c <E < 0$ (short dashed red), and $E> 0$ (dotted gray). The arrows show directions of the quantum state evolution along the paths. These most likely curves have the same form as in the plain measurement case, but with tunable collapse points. (b) Samples of numerically simulated qubit trajectories are plotted (solid gray fluctuating curves) along with the most likely paths for three different initial states: $\theta_I = 0.1 \pi$ (solid green), $\theta_I = 0.5 \pi$ (dashed purple), and $\theta_I = 1.2 \pi$ (dotted orange), and the collapse states are shown as thin dashed black horizontal lines. These most likely paths are computed numerically from Eqs.~\eqref{eq-fbodes}, using the initial states $\theta(t=0) = \theta_I$ and the final condition $p_{\theta}(T) = 0$ (the most likely paths without final fixed quantum states). For each set of boundary conditions (the 3 sets), multiple solutions are found. Both of the dotted orange curves represent local maximum where the bottom one is more likely than the other. For the solid purple set, both curves are equally likely. For the dashed green curve, the initial condition is close enough to an attractor that we only find one most likely solution. All curves in the three sets correspond to the stochastic energies $E_c < E < 0$.}
\label{fig-phasespace}
\end{figure*}

The optimal paths of the action in Eq.~\eqref{eq-fbaction} are obtained by extremizing the action over all variables $p_{\theta}$, $\theta$, and $r$, leading to two ordinary differential equations and one constraint,
\begin{subequations}\label{eq-fbodes}
\begin{align}
\label{eq-fbodesa}\dot{\theta} =&\, \Delta_0 + \Delta_1\, r - r\sin \theta /\tau_m,\\
\dot{p_{\theta}} = & \,p_{\theta}\, r\cos\theta  /\tau_m + r \sin\theta  /\tau_m ,\\
\label{eq-fbodes-c}r = &\, p_{\theta} \Delta_1 \tau_m - p_{\theta} \sin\theta + \cos\theta,
\end{align}
\end{subequations}
with arbitrary boundary conditions on the variable $\theta$ and $p_{\theta}$. The solutions of these equations are the most likely paths of the system. By writing the action in the form ${\cal S}_{fb} = \int \!\! \dd t(- p_{\theta} \dot{\theta} + {\cal H}[p_{\theta}, \theta, r])$, where ${\cal H} =  p_{\theta}(  \Delta_0 +\Delta_1 \, r - r \sin\theta/\tau_m)- (r^2 - 2 \,r \cos \theta +1)/2 \tau_m$ is a \textit{stochastic Hamiltonian}, we can examine the most likely paths by using phase space analysis~\cite{Chantasri2013}, motivated by the phase space concept in classical mechanics. Substituting the $r$ constraint in Eqs.~\eqref{eq-fbodes-c} into the action (equivalent to integrating the path integral over the variable $r(t)$), we obtain the action and the stochastic Hamiltonian in terms of only the system variable $\theta$ and its conjugate $p_{\theta}$,
\begin{align}\label{eq-fbstoHam}
{\cal H} = \frac{(p_{\theta} \Delta_1 \tau_m + \cos \theta - p_{\theta} \sin\theta)^2}{2 \tau_m} + p_{\theta} \Delta_0 - \frac{1}{2\tau_m}.
\end{align}
This quantity is explicitly time-independent and so it is a constant of motion for the most likely path, a solution of the ordinary differential equations in Eqs.~\eqref{eq-fbodes}. Let us define the stochastic energy $E = {\cal H}$, and then solve for the conjugate variable $p_{\theta}(\theta, E)$ as a function of $\theta$ and $E$ from Eq.~\eqref{eq-fbstoHam}. Each value of $E$ will then correspond to a curve in the phase space portrait, describing dynamics of the most likely paths, the same way the individual classical trajectories are depicted as constant-energy curves on a phase space plot.

As an example, we show in Figure~\ref{fig-phasespace}(a) the phase space portrait for the pure linear feedback case (i.e., $\Delta_0 = 0$), where we plot the conjugate variable $p_{\theta}$ as a function of $\theta$,
\begin{align}\label{eq-fbconjg}
p_{\theta}(\theta, E) = \frac{-\cos\theta \pm \sqrt{1+2 E \tau_m}}{\Delta_1 \tau_m-\sin\theta},
\end{align}
for different values of $E$. This is an interesting case. We can see from the phase space plot in Figure~\ref{fig-phasespace}(a) that there exists some attractors to which all states eventually limit to.

These attractors coincide with the divergence of the conjugate variable $p_{\theta}(\theta, E)$ in Eq.~\eqref{eq-fbconjg}, and also appear as stationary points where $\dot{\theta} = 0$ in Eqs.~\eqref{eq-fbodesa}. Solving these equations, we obtain the attractors are located at $\theta_{s1} = j \pi + \arcsin( \Delta_1 \tau_m)$ and $\theta_{s2} = (j+1)\pi - \arcsin(\Delta_1 \tau_m)$ where $j = 0,2,4,...$ is an even integer. These attractors may be interpreted as stabilized states achieved by turning on a linear feedback $0< |\Delta_1 \tau_m | \le 1$ and turning off the bare qubit frequency $\Delta_0 =0$. They are effectively the new collapse points, rather than at the poles of the Bloch sphere, i.e., $\theta = 0, \pi, 2 \pi,...$ ($z = \pm 1$). Considering only angles between $0$ and $2 \pi$, for the positive feedback $ 0 <\Delta_1 \tau_m \le 1$, as $|\Delta_1 \tau_m|$ increases from zero, the stabilized states $\theta_{s1}, \theta_{s2}$ move from $0$ and $\pi$ toward each other and coalesce at $\theta_{s1}= \theta_{s2} = \pi/2$ when $\Delta_1 \tau_m = 1$; whereas in the negative feedback $-1 \le \Delta_1 \tau_m < 0$, as $|\Delta_1 \tau_m|$ grows, the stabilized states move toward each other in the region of $\pi$ and $2\pi$ and coalesce at $3 \pi/2$ when $\Delta_1 \tau_m =- 1$. We simulate numerical qubit trajectories using the Monte Carlo method (Appendix~\ref{app-numer}), showing that individual trajectories initialized at different states are eventually pinned to the stabilized states as predicted from the most likely path phase space. This is shown in Figure~\ref{fig-phasespace}(b) for three different initial conditions, along with their most likely paths.

\subsection{Stabilizing Rabi oscillation and its correlation function}
In this subsection, we show an example using the path integral to compute a correlation function for a system with linear feedback loop. The feedback loop of this example resembles the one used in the solid-state qubit measurement in Ref.~\cite{Korotkov2005-2} which has been adapted and realized in the transmon qubit experiment~\cite{Korotkov2012}. The theoretical setup is an oscillating qubit (a double quantum dot) continuously monitored by a near quantum-limited detector (a quantum point contact), of which the Hamiltonian and the measurement operator are in the same form as presented in Section~\ref{sec-qubit}. We as before consider the ideal symmetric qubit case where $\epsilon = 0$, and the qubit is measured with an efficient detector with no extra environment dephasing. The qubit state is represented by a single parameter $\theta$ where $z=\cos \theta$ and $y= \sin \theta$. The stochastic master equation for the qubit state, in the $\delta t$-expansion similar to Eqs.~\eqref{eq-contstate} (Stratonovich form, as discussed in Ref.~\cite{Korotkov2005-2}), is given by,
\begin{align}\label{eq-fbmaster}
\dot{\theta} = \Delta_{fb}(\theta) - \sin \theta \cos \theta/\tau_m- \xi\,\sin \theta/\tau_m^{1/2},
\end{align}
where $\theta$ and $\xi$ are functions of time and $\Delta_{fb}$ is the feedback Rabi frequency, which is assumed to be a function of the qubit state $\theta$. We later neglect the second term on the right side of Eq.~\eqref{eq-fbmaster} because we consider a diffusive Rabi limit $\Delta \gg \tau_m^{-1}$.

The feedback protocol considered in this subsection (and also in Ref~\cite{Korotkov2005-2}) is a linear feedback designed to stabilize the quantum oscillation of the qubit state, against the random phase kicks due to the measurement. The desired qubit evolution is described by $y = \sin( \Delta_d \,t)$ and $z=\cos(\Delta_d \,t)$ where we define $\Delta_d$ as the target oscillation frequency. The difference between the actual phase $\theta$ and the target phase $\Delta_d \,t$, denoted as $\delta \theta(t) = \theta(t) - \Delta_d\, t$, is used to control the oscillating part of the qubit Hamiltonian. Therefore, we write the feedback Rabi frequency as, $\Delta_{fb} = \Delta_d(1- F \,\delta \theta)$, where $F$ is the dimensionless feedback factor. This feedback loop continuously corrects the random changes of the state made by the measurement so that the outcome trajectory closely follows the desired qubit oscillation.

Let us assume that the phase difference $\delta \theta$ is a slowly changing variable as compared to the oscillation with the desired frequency $\Delta_d$. Therefore, we can average the fluctuating process described in Eq.~\eqref{eq-fbmaster} over the oscillation period $\td{\delta t} = 2 \pi/\Delta_d$, taking $\delta \theta$ to be constant during this period. The period $\td{\delta t}$ will eventually be our new time scale. We define a new noise $\td{\xi}$ as a time-average of the last term of Eq.~\eqref{eq-fbmaster} over the oscillation period, $\td{\xi}(t) =  -(1/\td{\delta t})\int_t^{t+\td{\delta t}}\!\dd t'  \sin(\delta \theta(t') + \Delta_d\,t') \, \xi(t') /\tau_m^{1/2} $ with its zero ensemble average $\la \td{\xi}(t) \ra = 0$ and its variance being $\la \td{\xi}(t)^2 \ra =(1/\td{\delta t})^2 \int_t^{t+\td{\delta t}}\! \dd t' \, \sin^2(\delta \theta(t') + \Delta_d \, t') /\tau_m = 1/2 \tau_m \td{\delta t}$. We then can simplify the differential equation Eq.~\eqref{eq-fbmaster} to one in terms of the phase difference $\delta \theta$,
\begin{align}\label{eq-dthetamaster}
\dot{\delta \theta}(t) =- F \Delta_d \delta \theta(t) +  \td{\xi}(t),
\end{align}
with the new time scale $\td{\delta t}$.

Here we will use the above differential equation Eq.~\eqref{eq-dthetamaster} to compute a correlation function $K_z(\tau)= \la z(t) z(t+\tau)\ra$ using the path integral approach. Following the derivation of the action in Eq.~\eqref{eq-actiongenS} and \eqref{eq-contaction} where $\qq = \delta \theta$ and $P(\td{\xi}) = (\tau_m \td{\delta t}  / \pi)^{1/2} \exp(- \tau_m \td{\xi}^2 \td{\delta t})$ are now our new system variable and a noise probability density function, we obtain the action in this form,
\begin{align}
{\cal S}_{fb} = \int_0^T\!\!\! \dd t' \big\{- i p_{\delta \theta}(\dot{\delta \theta} + F \Delta_0 \delta \theta -\td{\xi}) - \tau_m \td{\xi}^2 \big\},
\end{align}
omitting the boundary terms. Note that we have written the pure imaginary conjugate variable ($i p_{\delta \theta}$) explicitly with $i$. As before, we can compute an average quantity by integrating the paths $\la \cdots \ra = {\cal N}\!\! \int \!\! {\cal D}\delta \theta\, {\cal D}p_{\delta \theta} {\cal D} \td{\xi} e^{{\cal S}_{fb}} \cdots$. So, we first write the correlation function $K_z(\tau)$ in terms of the phase difference $\delta \theta$, then average over the oscillation period $\td{\delta t}$, getting rid of the fast fluctuating parts, and we are left with $K_z(\tau) \approx \la \cos[\delta \theta(t)-\delta \theta(t+\tau)]\ra \cos( \Delta_d \tau)/2 + \sin[\delta \theta(t)-\delta \theta(t+\tau)]\ra \sin (\Delta_d \tau)/2$. The correlation function apparently can be written in terms of the real part and imaginary part of the following quantity, 
\begin{align}\label{eq-fbintegral}
\la e^{i \delta \theta(t) - i\delta \theta(t+\tau)}\ra = {\cal N}\!\!\! \int \!\!\! {\cal D}\delta \theta\, {\cal D}p_{\delta \theta} {\cal D} \td{\xi} e^{\cal S} e^{i \delta \theta(t) - i \delta \theta(t+\tau)},
\end{align}
where we have used the same notations, such as $\int \!\! {\cal D}\delta \theta$ for the integral over all possible paths $\delta \theta(t)$, and ${\cal N}$ for a normalized factor.

The Gaussian integral over $\td{\xi}$ in Eq.~\eqref{eq-fbintegral} is quite straightforward resulting in a bilinear term in $p_{\delta \theta}$, which then leads to another Gaussian integral of $p_{\delta \theta}$. As one would expect, these integrals generate another prefactor that cancels the normalized factor ${\cal N}$. Consequently, the last integral over $\delta \theta$ is left as $\la e^{i \delta \theta(t) - i \delta \theta(t+\tau)}\ra  = \int\!\!{\cal D}\delta \theta \,e^{{\cal S}'}$ where the exponent ${\cal S}'$ is given by,
\begin{align}
\nn{\cal S}'= \int_0^T\!\! \dd t' \big\{ & -\tau_m(\dot{\delta \theta}+ F \Delta_d\, \delta \theta)^2 \\
& + i \, \delta \theta\, \delta(t'-t) - i \,\delta \theta\, \delta(t' - t- \tau)\big\}.
\end{align}
This effective action can be transformed further using the integration by parts, for example, $\int \!\! \dd t \,\dot{\delta \theta}\dot{\delta \theta} =- \int\!\! \dd t\, \delta \theta\, \ddot{\delta \theta}$, assuming that the boundary conditions for $\delta \theta$ at both end points vanish. We then write the action in terms of the inverse Green's function and the source term as discussed in section~\ref{sec-briefintro}, ${\cal S}' = - \frac{1}{2} \int\!\! \dd t' \dd t'' \delta \theta(t') G_{\delta \theta}^{-1}(t',t'') \delta \theta(t'') + \int \!\!\dd t' J_{\delta \theta}(t') \delta \theta(t')$ where the inverse Green's function, the Green's function, and a particular form of the source term are given by,
\begin{align}
G_{\delta \theta}^{-1}(t',t'') &=  2 \tau_m \delta(t'-t'')\left( -\partial_{t''}^2 + F^2 \Delta_d^2 \right),\\
G_{\delta \theta}(t',t'') &= 1/(4 \tau_m | F \Delta_d|)\exp\left\{- |F \Delta_d (t' - t'') |\right\},\\
J_{\delta \theta}(t') &= i \,\delta(t'-t) - i\, \delta(t'-t-\tau).
\end{align}

Performing the last Gaussian functional integral over $\delta \theta$ gives,
\begin{align}
\nn\la e^{i \delta \theta(t) - i \delta \theta(t+\tau)}\ra  =& \exp\left\{\frac{1}{2} \int \!\! \dd t' \dd t'' J(t') G(t', t'') J(t'')\right\}\\
 =&  \exp\left\{\frac{e^{- |F \Delta_d| \tau}-1}{4 \tau_m |F \Delta_d|}\right\},
\end{align}
for the positive value of the time difference $\tau$. This quantity is a real number, therefore we obtain the correlation function,
\begin{align}
 K_z(\tau) = \frac{\cos (\Delta_d \tau)}{2} \exp\left\{\frac{e^{- F \Delta_d\, \tau}-1}{4 \tau_m F \Delta_d}\right\},
\end{align}
assuming that $F \Delta_d \ge 0$. This result agrees with the solution found in Ref.~\cite{Ruskov2002} which is presented in different notation.

\section{Conclusion}\label{sec-conclusion}
We have developed and extended the stochastic path integral technique to study statistical behaviour of a quantum system under weak continuous measurement, as well as measurement with feedback, presented with several qubit examples. The path integral approach is constructed based on the joint probability distribution of the measurement records, which is then extended to the distribution of quantum states, describing all possible quantum trajectories. We have shown that with this path integral and its action formalism, the optimal dynamics, such as the most likely paths, can be obtained naturally from the extremization of the action, whereas other statistical quantities can be achieved from direct integration or perturbation theory. In the case of plain measurement of a qubit, we have derived analytic solutions for the average trajectory, the variance, and the correlation functions conditioning on the fixed initial and final states, which show an excellent agreement with the numerically simulated data. 

We have also presented a diagrammatic perturbation method used in computing expectation values and correlation functions of quantum trajectories, and elaborated it with examples of the qubit with Rabi oscillation case. The variances and multi-time correlation functions of qubit trajectories in the short-time regime have been revealed using this method given initial conditions, and the results are in good agreement with the numerical simulation. Moreover, we have considered quantum measurement with feedback control, using the action principle to investigate the dynamics of the most likely paths of a qubit with linear feedback on its oscillating frequency. We have discovered that the direct linear feedback, manipulating the qubit Hamiltonian instantly using the measurement readout, can stabilize the qubit state to arbitrarily chosen pure states. We have also considered the example of the feedback loop stabilizing the qubit Rabi frequency introduced in Ref.~\cite{Korotkov2005-2}, and we have computed the correlation function for the qubit trajectory using the path integration method.

So far, the stochastic path integral formalism in the context of continuous quantum measurement has been proven to be useful in studying the statistics of quantum trajectories; however, there are some unsolved issues that need to be further explored. One is the limitation of the statistical average solutions derived from the perturbation expansion theory. Only the first few orders of the expansion have been computed, resulting in the solutions that are valid only in the certain parameter regimes. We hope to find solutions in an arbitrary regime, possibly with some modifications of our approach. Another issue is the assumption of the instantaneous feedback, which can be difficult to realize in experiments. Feedback loops modelled with time delays will be taken into account in future work.

\begin{acknowledgments}
We thank J. Dressel for discussions, helpful comments on the manuscript, and for collaborating in our first paper on this subject \cite{Chantasri2013}. We thank I. Siddiqi and S. G. Rajeev for discussions. This work was supported by US Army Research Office Grants No. W911NF-09-0-01417 and No. W911NF-15-1-0496, by National Science Foundation grant DMR-1506081, by John Templeton Foundation grant ID 58558, and by Development and Promotion of Science and Technology Talents Project Thailand.
\end{acknowledgments}

\appendix
\section{Connection to other path integral formalisms for quantum measurement}\label{app-otherpathint}
There are interesting comparisons between the path integral formalism we introduced in the main text and other path integral approaches built upon Feynman path integral in quantum mechanics \cite{BookFeyHib}. As we have shown, our formalism is aimed at describing the probability distribution of quantum trajectories, paths of state of a system under continuous measurement, on its quantum state space (such as Hilbert space for pure states). Each individual quantum trajectory in the path integral is realizable and tractable, as demonstrated experimentally, such as, in solid-state systems \cite{Kater2013,Weber2014}. However, for other path integral formalisms developed from the Feynman path integral to investigate quantum systems under measurement \cite{Feyn1948, Mensky1979, *Mensky1994,Caves1986, *Caves1987,Barchielli1982,Presilla1996}, the evolution of the quantum state (or wavefunction) is based on interference of all possible classical paths in measurable configuration space (such as the system's positions or spin states). Thus, in this latter case, the integrations are over configuration coordinates of the measured system.

Despite the differences in forms, the two approaches mentioned above can be related. The path integral via Feynman's concept can be used to compute the probability distribution of the measurement results, which is then, as shown in the main text, one of the most important ingredients in constructing our stochastic paths in quantum state space. To elaborate this connection in more detail, we consider the path integral method presented by Caves \cite{Caves1986,*Caves1987}, for measurements providing information about the position $x(t)$ of a nonrelativistic, one-dimensional quantum system, evolved in time. In that approach, the effect of measurements is to restrict the sum over paths by weighting each path differently depending on the measurement results. These weights appear in the path integral in Ref.~\cite{Caves1986,*Caves1987} as the `resolution amplitude', $\Upsilon({\bar x}(t) - x(t))$, which also accounts for the imprecision of the measurements. 

Let us assume instantaneous position measurements that are equally distributed in times, at $t_0, t_1, ..., t_{n-1}$ where $t_k = t_0 + k \delta t$, giving measurement readouts denoted by ${\bar x}_0, {\bar x}_1, ..., {\bar x}_{n-1}$. The joint probability amplitude $\Phi$ of the measurement records, and that the system is at $x_n$ at time $t_n$, given an initial wavefunction $\psi_0$, is in this form,
\begin{align}\label{eq-cavePI}
\nn\Phi({\bar x}_0,...,{\bar x}_{n-1}; &x_n, t_n  |\psi_0 ) = \int\!\!\! {\cal D}x(t)\left( \prod_{k=0}^{n-1} \Upsilon({\bar x}_k - x_k)\right)\\
 &\times \left( \prod_{k=0}^{n-1} \la x_{k+1} | e^{-i \delta t {\op H}} | x_k \ra \right) \la x_0 | \psi_0 \ra,
\end{align}
where the position coordinates are denoted by $x_0, x_1, ..., x_n$ at time $t_0, t_1,...,t_n$, and the integral measure is defined as $\int\!\! {\cal D}x(t) \equiv  \int \dd x_0 \cdots \dd x_{n-1}$. We note here that we have modified the ordering of the measurements from the original version by Caves, in which he assumed that the measurements start only after the initial state has evolved for time $\delta t$. We assume that this change has an infinitesimal effect on the probability amplitude as $\delta t \rightarrow 0$. The first bracketed term on the right hand side of Eq.~\eqref{eq-cavePI} describes the influence of the measurement. Without it, the path integral will be the usual Feynman path integral, exactly equal to $\la x_n | e^{-i T {\op H}}| \psi_0\ra$ or the wave function at the final time $t_n = T$. 

From Eq.~\eqref{eq-cavePI}, the joint probability distribution function (PDF) of the sequence of measurement readouts can be obtained by integrating the square of the probability amplitude over the final coordinate $x_n$,
\begin{align}\label{eq-cavejpdf}
P({\bar x}_0, {\bar x}_1 , ... , {\bar x}_{n-1}|\psi_0) =\!\! \int \!\!\!\dd x_n |\Phi({\bar x}_0,...,{\bar x}_{n-1};x_n, t_n |\psi_0 ) |^2,
\end{align}
given the initial wavefunction $\psi_0$.

In order to see that this joint PDF in Eq.~\eqref{eq-cavejpdf} is the same as what we have earlier in the main text (in this special case), we first rewrite this term $\int \!\! \dd x_0 \la x_1 | e^{-i \delta t {\op H}}\Upsilon({\bar x}_0 - x_0) | x_0 \ra \la x_0 |\psi_0\ra = \la x_1 | {\op U}_{\delta t} {\op M}_0 |\psi_0 \ra$ as a probability amplitude right after one measurement (${\op M}_0$) and one unitary transformation (${\op U}_{\delta t}$). Then, we obtain a probability density function for the first measurement outcome as $P({\bar x}_0 | \psi_0) = \int \!\! \dd x_1 |  \la x_1 | {\op U}_{\delta t} {\op M}_0 |\psi_0 \ra |^2$. We further introduce an updated wavefunction $|\psi_{k+1}\ra$, a result of a state $|\psi_k \ra$ that has gone through one measurement ${\op M}_k$ and one unitary transformation ${\op U}_{\delta t}$,
\begin{align}\label{eq-caveupdate}
\nn|\psi_{k+1} \ra & = \frac{\int \dd x_{k} e^{-i \delta t {\op H}}\Upsilon({\bar x}_{k} - x_{k})|x_{k} \ra \la x_{k} | \psi_{k}\ra}{\sqrt{P({\bar x}_{k} | \psi_{k})}},\\
&= \frac{{\op U}_{\delta t} {\op M}_k | \psi_k \ra}{\sqrt{P({\bar x}_{k} | \psi_{k})}},
\end{align}
where the denominator assures the correct norm of the wavefunction. Then, the joint distribution of the measurement readouts of the first two times is given by $P({\bar x}_1, {\bar x}_0 |\psi_0) = \int \!\! \dd x_2 | \la x_2 | {\op U}_{\delta t} {\op M}_1| \psi_1 \ra |^2 P({\bar x}_0 | \psi_0) = P({\bar x}_1 | \psi_1)P({\bar x}_0 | \psi_0)$. Repeating this procedure further to the next measurement readouts ${\bar x}_2, {\bar x}_3, ... {\bar x}_{n-1}$, at the end, we obtain the full joint PDF,
\begin{align}\label{eq-cavejpdf2}
P({\bar x}_0, {\bar x}_1 , ... , {\bar x}_{n-1}|\psi_0) = \prod_{k=0}^{n-1} P({\bar x}_k | \psi_k),
\end{align}
as a product of the conditional probability density functions, defined in a general form as $P({\bar x}_k | \psi_k) =  \int \!\! \dd x_{k+1} | \la x_{k+1} | {\op U}_{\delta t} {\op M}_k| \psi_k \ra |^2$. This is the joint PDF of the measurement readouts, the main component of the joint PDF in Eq.~\eqref{eq-mainpdf}.

To obtain an analogous form of the joint PDF in Eq.~\eqref{eq-mainpdf}, we add probability density functions of the quantum state (or wavefunction) as delta functions to every single time step. We obtain the full joint PDF,
\begin{align}
{\cal P}_{\psi} = \prod_{k=0}^{n-1}P(\psi_{k+1} | \psi_k, {\bar x}_k)P({\bar x}_k |\psi_k),
\end{align}
where the update state in this case is written as $P(\psi_{k+1} | \psi_k, {\bar x}_k) = \delta^d\big\{|\psi_{k+1}\ra -  {\op U}_{\delta t} {\op M}_k | \psi_k \ra/\sqrt{P({\bar x}_{k} | \psi_{k})}\big\}$ as in Eq.~\eqref{eq-caveupdate}. The integer $d$ is the dimension of the vector $|\psi_k\rangle$ (which maybe generalized to infinite dimension via a functional form of the $\delta$-function), and ${\cal P}_{\psi}$ is called a joint probability density function of the measurement outcomes $\{ {\bar x}_k \}$ and the wavefunctions $\{ | \psi_k \ra \}$ of the measured system.

We note that our path integral presented in the main text deals with mixed states instead of pure states, and the system we consider is the qubit (or spin) system with the discrete basis, i.e., $\int \dd x |x\ra \la x| \Rightarrow \sum_{s=1,-1} |s\ra \la s|$. Therefore, the operator ${\op M}_k$ in the discussion above is equivalent to what we have as the measurement operator ${\op M}_k = {\op M}_{r_k}= \Upsilon(r_k-1)|1\ra \la 1| + \Upsilon(r_k +1) |-1 \ra \la -1| $, defined in Section~\ref{sec-qubit}, and the resolution function is a Gaussian function, $\Upsilon(r_k - s) = (\delta t/2 \pi \tau_m)^{1/4}\exp\{-\delta t(r_k - s)^2/4 \tau_m\}$ for $s =1$ and $s=-1$.

It is also worth mentioning that in our approach, the stochastic trajectories in quantum state spaces can be thought of as classical trajectories in configuration space, such as trajectories on the Bloch sphere can be considered as three-dimensional random walks in a unit-radius sphere. Then, one could find connections between our path integral and the formalism in classical stochastic processes, such as the Pilgram-Sukhorukov-Jordan-B{\"u}ttiker path integral \cite{Jordan2003,*Jordan2004,*JordanSuk2004,*SukhorJordan2007}, the Martin-Siggia-Rose formalism, the Wiener integral, or Feynman-Kac path integral (see Eqs.~\eqref{eq-probzz},\eqref{eq-probzz2}, and \eqref{eq-derivep}), however, we do not cover the discussion in this paper.

\section{The path integral's action for the qubit measurement}\label{app-actionpure}
We show the derivation here, how the action in Eq.~\eqref{eq-probzz2} can be written in this form,
\begin{align}
{\cal S}[u] = {\cal S}[\bar{u} ] - \frac{\tau_m}{2 \delta t} \sum_{k=0}^{n-1}(\eta_{k+1}-\eta_k)^2,
\end{align}
where ${\bar u}$ is the optimal path, extremizing the action ${\cal S}[u]$.

To see this, one can Taylor expand the action Eq.~\eqref{eq-probzz2} in discrete forms, such as ${\cal S}[u] = {\cal S}[\bar{u} +\eta] = {\cal S}[\bar{u}] + {\cal S}'[{\bar u}]\eta + \frac{1}{2}{\cal S}'[{\bar u}]\eta^2 + O(\eta^3)$, and show that higher order terms  vanish ($O(\eta^3) = 0$). However, there is a simpler way to see this, using the action in a continuous form,
\begin{align}\label{eq-appcontS}
{\cal S} = - \int_0^T \!\!\! \dd t \bigg \{\frac{ \tau_m}{2}{\dot u}(t)^2 -\tanh u(t) {\dot u}(t)+ \frac{1}{2 \tau_m}\bigg\},
\end{align}
which is the same action as in Eq.~\eqref{eq-probzz2}, with definitions of the time integral, $\delta t \sum_{k=0}^{n-1} A_k = \int_0^T\!\! \dd t A(t)$, and the derivative, $(A_{k+1} - A_k)/\delta t = {\dot A}(t)$. In this form, one can see that the last two terms in the action can be integrated easily. The second term can be written in this general form,
\begin{align}
\nn \int_0^T \!\!\! \dd t \, F'(u(t)) {\dot u}(t) &= \int_0^T \!\!\! \dd t \, {\dot F}(u(t)), \\
&= F(u(T)) - F(u(0)),
\end{align}
where $F(u(t)) = \int \!\! \dd u \tanh u = \ln(\cosh u(t))$. The contribution of this term to the action is only dependent on the boundary terms. Therefore, we can write the last two terms in Eq.~\eqref{eq-appcontS} as equivalent to $\int_0^T \!\! \dd t \left\{ \tanh {\bar u}(t) {\dot {\bar u}}(t) - 1/ 2 \tau_m \right\}$. 

After this simplification, the first term in Eq.~\eqref{eq-appcontS} can be written as $ - \int_0^T \!\! \dd t \frac{\tau_m}{2}\left\{ {\dot {\bar u}}(t)^2 + 2 {\dot {\bar u}}(t){\dot \eta}(t) + {\dot \eta}(t)^2\right\}$, performing the integration by parts giving $\int_0^T \!\! \dd t\, {\dot {\bar u}}(t) {\dot \eta}(t) = - \int_0^T \!\! \dd t \,{\ddot {\bar u}}(t)\eta(t)=0$. The action is then given by,
\begin{align}
{\cal S}[u] = {\cal S}[{\bar u}] - \frac{\tau_m}{2}\!\!\! \int \!\!\! \dd t \,{\dot \eta}(t)^2,
\end{align}
where,
\begin{align}
{\cal S}[{\bar u}] =-  \int_0^T \!\!\!\dd t \left\{ \frac{\tau_m}{2} {\dot {\bar u}}(t)^2 -  \tanh {\bar u}(t) {\dot {\bar u}}(t) + \frac{1}{2 \tau_m} \right\},
\end{align}
is the action in terms of the optimal solution ${\bar u}(t)$.

\section{Inverse matrix ${\bm M}^{-1}$}\label{app-invmatrix}
We show here the inverse of the matrix ${\bm M}$ in Eq.~\eqref{eq-matrixB}, assuming that it is a square matrix with $d$ dimensions. The inverse matrix ${\bm M}^{-1}$ is given by,
\begin{align} \nn
{\bm M}^{-1}=\frac{\left(\frac{\delta t}{\tau_m}\right)}{d+1} \left ( \begin{matrix} d & d-1 & d-2 & \cdots & 1\\ d-1 & 2(d-1) & 2(d-2) & \cdots & 2(d-d+1) \\ d-2& 2(d-2) & 3(d-2) & \cdots & 3 \\ \vdots & \vdots & \vdots & \ddots & \vdots \\ 1 & 2 & 3 & \cdots & d \end{matrix} \right),
\end{align}
where the diagonal elements are of the form $M^{-1}_{kk} = (\delta t/\tau_m)k(d-(k-1))/(d+1) =(\delta t/\tau_m)k(n-k)/n$, where $d = n-1$, and the other off-diagonal elements are $M^{-1}_{jk} = M^{-1}_{kj}=(\delta t/\tau_m)j(n-k)/n$ for $k\ge j$ as verified by direct calculation.

\section{Full solutions for the qubit measurement without Rabi oscillation}\label{app-fullqnd}
The solutions of the preselected and postselected average and variance in $z$ to infinite order are shown here. The average is given in terms of the infinite summation as,
\begin{align}
_{z_F}\langle z_j \rangle_{z_I}= \sum_{m=0}^{\infty} \frac{1}{(2 m)!}\left(\frac{d^{2m}}{d u_j^{2m}}\tanh(u_j)\bigg|_{u_j = \bar{u}_j}\right)\langle \eta_j^{2m} \rangle,
\end{align}
where the differentials in the bracket are evaluated at the optimal path ${\bar u}$. Similarly, the variance is given by,
\begin{align}
\nn _{z_F}\langle z_j^2 \rangle_{z_I} - &\,_{z_F}\langle z_j \rangle_{z_I}^2\\
\nn&= \sum_{m=0}^{\infty} \frac{1}{(2 m)!}\left(\frac{d^{2m}}{d u_j^{2m}}\tanh^2(u_j)\bigg|_{u_j = \bar{u}_j}\right)\langle \eta_j^{2m} \rangle\\ 
& - \left(\sum_{m=0}^{\infty} \frac{1}{(2 m)!}\left(\frac{d^{2m}}{d u_j^{2m}}\tanh(u_j)\bigg|_{u_j = \bar{u}_j}\right)\langle \eta_j^{2m} \rangle\right)^2.
\end{align}
The average $\la \eta_j^{2 m} \ra$ are given explicitly in Eq.~\eqref{eq-etamo} in the main text.

\section{Numerical simulation}\label{app-numer}
The numerical data presented in Figure~\ref{fig-qnd}, \ref{fig-corrlimit} and \ref{fig-phasespace} are from the simulation of quantum trajectories using Monte Carlo method. The numerical trajectories are generated in $n$-discrete steps of $\delta t$. Starting from an initial state $\qq_0$, each step, we randomly generate a measurement readout from a distribution, for example, $P(r_k |\qq_k)$, and compute a quantum state from the update equation $\qq_{k+1} = {\bm {\mathcal E}}[\qq_k, r_k]$ (or $\rho_{k+1} = {\cal O}_{\gamma} {\cal U}_{\delta t} {\cal M}_{r_k}[\rho_k]$), repeating the procedures from $k =0$ to $n-1$ to get a full trajectory $\{ \qq_k\}_0^n$. 

In Figure~\ref{fig-qnd}, we simulate the data using $n = 100$ time steps, with the step size $\delta t =  0.006$ and $\tau_m = 1$. In Figure~\ref{fig-corrlimit}, we use $n=500$, $\delta t = 0.01$, and $\tau_m = 10$, where the Rabi frequency $\Delta$ is set to $2\pi$. We note that in the main text we present these numbers in the unit of $\tau_m$.

For the data presented in Figure~\ref{fig-phasespace}, where the Rabi oscillation is linearly dependent on the highly fluctuating measurement readouts, the state update equation needs to be modified to minimize numerical errors in each time step of the calculation. The update state is then computed from $ \rho_{k+1} = {\cal U}_{\delta t}^{1/10} {\cal M}_{r_k}^{1/10} \cdots \,\, {\cal U}_{\delta t}^{1/10} {\cal M}_{r_k}^{1/10}[\rho_k]$, where the measurement operator ${\cal M}_{r_k}$ and unitary operator ${\cal U}_{\delta t}$ are each divided into 10 pieces and are operated onto the qubit state $\rho_k$ alternately. We simulate the data using $n=1000$ time steps, with the step size $\delta t = 0.01$ and $\tau_m = 1$. The feedback Rabi oscillation is $\Delta(t_k) = 0.8\, r_k$, where $r_k$ is a generated measurement readout at time $t_k$.

\section{White noise limit}\label{app-whitenoise}
In the main text, the state updating procedure is based on the assumption that a measurement outcome is randomly generated from a readout distribution $P(r_k|\qq_k)$, which is a function of a state right before the measurement. However, in the limit when the measurement readout is highly fluctuating, for example, the standard deviation of the readout distribution is much larger than the measurement response, $(\tau_m/\delta t)^{1/2} \gg 1$, we can approximate the readout as a sum of two parts, one being its average (which is related to the measured system state) and another being a zero-mean independent fluctuating noise. Following the readout distribution in Eq.~\eqref{eq-probr}, the average of the readout at a time $t_k$ is exactly $\langle r_k \rangle = z_k$, therefore, we write $r_k= z_k + \eta_k$ where $\eta_k$ is a zero-mean Gaussian white noise.

The next step is to find a probability distribution for the independent noise $\eta_k$, which we then approximate as having the same variance as the original distribution, which in this case the variance is $\tau_m/\delta t$. This approximation is valid as long as the variance is much larger than the separation between two outcomes $r = 1$ and $r =-1$. A more rigorous proof can be done by writing $\eta_k = \sqrt{\tau_m} \xi_k$ where $\xi$ is a Gaussian white noise with variance $\delta t^{-1}$. Note that $\xi$ is the time-derivative of the Wiener increment, scaling as $\delta t^{-1/2}$  (It\^{o} calculus) \cite{BookGardiner}, therefore, in any expansion, we need to keep terms containing $r_k^2 \delta t^2 \approx \tau_m \delta t$. Using these rules, we can write,
\begin{align}
\nn P(r_k | z_k) &= {\cal N} \left(\frac{1+z_k}{2}e^{-\frac{\delta t}{2 \tau_m}(r_k-1)^2 } +\frac{1-z_k}{2}e^{-\frac{\delta t}{2 \tau_m}(r_k+1)^2 }\right) \\
&\approx {\cal N}\exp\left\{-\frac{\delta t}{2 \tau_m}(r_k-z_k)^2\right\},\label{eqcompareP}
\end{align}
where these two equations are exactly equal in the expansion up to the first order of $\delta t$. The second line confirms that the distribution of $\eta_k$ is Gaussian with the same variance $\tau_m/\delta t$ and the mean $z_k$.

\section{The initial-boundary source terms}\label{app-sourceterm}
Let us assume that $x$ is the only system variable we consider, where an update equation is written as $x_{i+1} = x_i + L(x_i, \xi_i)\delta t$ and the probability density function for a measurement readout is proportional to $e^{-\xi_i^2 \delta t/2}$. We then write the action for this system as,
\begin{align}
{\cal S} =& \sum_{i=0}^{n-1}\left\{ - p_i (x_{i+1} - x_i) + p_i L(x_i, \xi_i)\delta t - \frac{1}{2}\xi_i^2 \delta t\right\}.
\end{align}
where $p$ is the conjugate variable. In this case, we could write the first term, $\sum_{i=0}^{n-1} - p_{i} (x_{i+1} - x_i) = -\sum_{i=0}^{n-1} \sum_{j=0}^{n} p_i C_{i,j} x_j$, where $C_{i,j} = \delta_{i+1,j}-\delta_{i,j}$, however, the matrix $C$ would not be a square matrix, and its inverse is not simply defined. Therefore, we separate out one term, $p_0 x_0$, from the sum, which symmetrizes the double sum,
\begin{align}
\sum_{i=0}^{n-1} - p_{i} (x_{i+1} - x_i) = p_0 x_0 - \sum_{i=0}^{n-1}\sum_{j=1}^{n}  p_{i}(G^{-1}_x)_{i,j} x_j ,
\end{align}
resulting in a square matrix $(G^{-1}_x)$,
\begin{align}\label{eq-appGinvx}
(G^{-1}_x)_{i,j} =& \left( \begin{matrix} 1 & 0 & 0 & \cdots & 0 \\ -1 & 1 & 0 & \cdots & 0 \\ 0 & -1 & 1 & \cdots & 0 \\ \vdots & && \ddots & \vdots \\ 0 & \cdots & 0 & -1 & 1\end{matrix} \right)_{i,j},
\end{align}
with the row index, $i=0$ to $n-1$, and the column index, $j=1$ to $n$. For the measurement readout term, we define its square matrix,
\begin{align}\label{eq-appGinvxi}
(G^{-1}_{\xi})_{i,j} \equiv &\delta t\left( \begin{matrix} 1 & 0 & 0 & \cdots & 0 \\ 0 & 1 & 0 & \cdots & 0 \\ 0 & 0 & 1 & \cdots & 0 \\ \vdots & && \ddots & \vdots \\ 0 & \cdots & 0 & 0 & 1\end{matrix} \right)_{i,j},
\end{align}
with the row and column index being $i,j=0$ to $n-1$.

As a result, we can rewrite the action ${\cal S}$ into two separated terms, a free action ${\cal S}_F$ and an interaction action ${\cal S}_I$, where
\begin{align}
\label{eq-appfreeS}{\cal S}_F=&-  \sum_{i=0}^{n-1} \sum_{j=1}^{n} p_i (G^{-1}_x)_{i,j} x_j  - \frac{1}{2}\sum_{i=0}^{n-1}\sum_{j=0}^{n-1}\xi_i (G^{-1}_{\xi})_{i,j} \xi_j \\
{\cal S}_I =& \sum_{i=0}^{n-1} \bigg\{ p_0 x_0 \delta_{i,0}+ p_i L(x_i,\xi_i)\delta t \bigg\}.
\end{align}
The appearance of the first term in the interaction action, $p_0 x_0 \delta_{i,0}$, is the reason we have the extra terms (we represented them as $B$) in Eq.~\eqref{eq-interS}.

\section{The left continuous Heaviside step function}\label{app-heaviside}

We now show why the two-point correlation functions derived from our path integrals, such as the one derived from the free action in Eq.~\eqref{eq-appfreeS}, that is
\begin{align}
\langle x(t) p(t') \rangle_F = G_x(t,t') = \Theta(t-t'),
\end{align}
consist of a left continuous Heaviside step functions $\Theta(t)$ that behaves differently from the usual Heaviside step function. This left continuous Heaviside step function has the properties, $\Theta(0)=0$ and $\lim_{t\rightarrow 0^+} \Theta(t) = 1$, while the usual Heaviside step function has properties, $\Theta(0) = 1/2$, $\lim_{t\rightarrow 0^+} \Theta(t) = 1$, and $\lim_{t\rightarrow 0^-} \Theta(t) = 0$. This is true from a point of view of the discrete form of a correlation function, $\langle x_a p_b \rangle_F = (G_x)_{a,b}$, that the propagator has this property $(G_x)_{a,b} = 1$ when $a>b$ and it vanishes otherwise.

Let us start from writing a free generating function in a discrete form, $Z_F[J_x, J_p,J_\xi]$
\begin{widetext}
\begin{align}\label{eq-appgenz}
\nn Z_F[J_x, J_p, J_{\xi}]=& \frac{\left(\frac{\delta t}{2 \pi}\right)^{\frac{n}{2}}}{(2 \pi i)^{n}}\int \!\!\dd [x_k]_1^n \int \!\!\dd [p_k]_0^{n-1} \exp\left\{-\sum_{i=0}^{n-1}\sum_{j=1}^{n}  p_{i}(G^{-1}_x)_{i,j} x_j + \sum_{k=0}^{n-1} p_k (J_p)_k \delta t + \sum_{k=1}^{n} (J_x)_k x_k \delta t\right\}\\
&\times \int \dd [\xi_k]_0^{n-1} \exp\left\{- \frac{1}{2}\sum_{i=0}^{n-1}\sum_{j=0}^{n-1}\xi_i (G^{-1}_\xi)_{i,j} \xi_j   + \sum_{k=0}^{n-1} \xi_k (J_\xi)_k \delta t \right\}.
\end{align}
\end{widetext}

These matrix integrals can be carried out using multi-dimensional Gaussian integrals. The first one is,
\begin{align}
\int\! \!\dd p\, \dd x\, e^{-p^T\!\! A\,\, x + \, p^T \!J_p + J_x^T x} = e^{J_x^T A^{-1} J_p} \frac{(2 \pi i)^d}{\text{Det}[A]},
\end{align}
where $p$ is a pure imaginary $d$-dimensional vector, $x,J_p,J_x$, are real vectors, and $A$ is a real and invertible matrix. Another is,
\begin{align}
\int \!\!\dd \xi\, e^{-\frac{1}{2} \xi^T B \xi + \xi^T J_\xi} = e^{\frac{1}{2}J_u^T B^{-1} J_u} \frac{(2 \pi)^{d/2}}{(\text{Det}[B])^{1/2}},
\end{align}
where $\xi$ is a real $d$-dimensional vector and $B$ is a real and symmetric matrix. After integrating over all variables, the free generating function is given by,
\begin{align}
\nn Z_F=& \exp\left\{\sum_{i=1}^{n} \sum_{j=0}^{n-1} (J_x)_i (G_x)_{i,j} (J_p)_j \delta t^2\right\}   \\
& \times  \exp\left\{\frac{1}{2} \sum_{i=0}^{n-1} \sum_{j=0}^{n-1} (J_\xi)_i (G_\xi)_{i,j} (J_\xi)_j \delta t^2 \right\}.
\end{align}

We note that the integrals generate a prefactor $\frac{(2 \pi i)^n}{\text{Det}(G_x^{-1})} = (2 \pi i)^n$, knowing that $\text{Det}(G_x^{-1}) =1$, and another prefactor $ \left(\frac{2 \pi}{\delta t} \right)^{\frac{n}{2}}$, which both are then canceled with the prefactor in Eq.~\eqref{eq-appgenz}. Also, from Eqs.~\eqref{eq-appGinvx}-\eqref{eq-appGinvxi}, we can compute the inverses of them as,
\begin{align}\label{eq-appmatrixgx}
(G_x)_{i,j} = \left( \begin{matrix} 1 & 0 & 0 & \cdots & 0 \\ 1 & 1 & 0 & \cdots & 0 \\ 1 & 1 & 1 & \cdots & 0 \\ \vdots & && \ddots & \vdots \\ 1 & \cdots & 1 & 1 & 1\end{matrix} \right)_{i,j},
\end{align}
with the row index, $i=1$ to $n$, and the column index, $j=0$ to $n-1$, and
\begin{align}
(G_\xi)_{i,j} = \frac{1}{\delta t}\left( \begin{matrix} 1 & 0 & 0 & \cdots & 0 \\ 0 & 1 & 0 & \cdots & 0 \\ 0 & 0 & 1 & \cdots & 0 \\ \vdots & && \ddots & \vdots \\ 0 & \cdots & 0 & 0 & 1\end{matrix} \right)_{i,j}.
\end{align}
with the row and column indices, $i,j=0$ to $n-1$.

Now, let us compute the two-point correlation function $\langle x_a p_b \rangle_F$,
\begin{widetext}
\begin{align}\nonumber
\langle x_a p_b \rangle_F \equiv & \, {\cal N}\!\! \int \!\! \dd [x_k]_1^n \,\dd [p_k]_0^{n-1} \dd [\xi_k]_0^{n-1} (x_a p_b) \exp\left\{-\sum_{i=0}^{n-1}\sum_{j=1}^{n} p_{i}(G^{-1}_x)_{i,j} x_j - \frac{1}{2}\sum_{i=0}^{n-1}\sum_{j=0}^{n-1}\xi_i  (G^{-1}_\xi)_{i,j} \xi_j  \right\} \\ \nonumber
=&\frac{1}{\delta t}\frac{\partial}{\partial (J_x)_a}\frac{1}{\delta t} \frac{\partial}{\partial (J_p)_b} Z_F[J_x, J_p,J_\xi] \bigg |_{J_x,J_p,J_\xi=0} \\ \nonumber
=&  \frac{1}{\delta t}\frac{\partial}{\partial (J_x)_a} \frac{1}{\delta t} \frac{\partial}{\partial (J_p)_b}\exp\left[\sum_{i=1}^{n} \sum_{j=0}^{n-1} (J_x)_i (G_x)_{i ,j} (J_p)_j \delta t^2\right] \bigg |_{J_x,J_p=0}\\ 
=& (G_x)_{a,b} =  \begin{cases} 1 & \text{, if, } a \ge b+1 \text{, or, } a > b, \\ 0 & \text{, otherwise.} \end{cases}
\end{align}
\end{widetext}
where ${\cal N} = \left(\frac{\delta t}{2 \pi}\right)^{\frac{n}{2}}/(2 \pi i)^{n}$. The conclusion in the last line comes from the fact that the square matrix, $(G_x)$, has the row and column indices different by one time step (see Eq.~\eqref{eq-appmatrixgx}), which actually originates from the indices of, $\{ x_k \}_1^n$, and $\{ p_k \}_0^{n-1}$.

\section{The 10 diagrams of the correlation function $\langle v(t_1) v(t_2)\rangle$}\label{app-fullform}
Continued from Eq.~\eqref{eq-corrfourth}, we presented all 10 diagrams here,
\begin{widetext}
\begin{align}
\nn \langle \yp(t_1)& \yp(t_2) \rangle^{(0)}= \langle \yp(t_1)\yp(t_2)e^{{\cal S}_I}\rangle_F^{(0)},  \\ 
\nn =&
\,\,\begin{tikzpicture}[node distance=0.6cm and 0.8cm]
\coordinate[label=below:$v_{t_2}$] (b2);
\coordinate[right=of b2,label=below:$v_0$] (bp);
\coordinate[above=of b2,label=above:$v_{t_1}$](a1);
\coordinate[above=of bp,label=above:$v_0$](ap);
\draw[particle] (b2) -- (bp);
\draw[particle] (a1) -- (ap);
\draw (bp) circle (.06cm);
\draw (ap) circle (.06cm);
\fill[black] (b2) circle (.05cm);
\fill[black] (a1) circle (.05cm);
\end{tikzpicture}\,+
\begin{tikzpicture}[node distance=0.6cm and 0.8cm]
\coordinate[label=below:$v_{t_2}$] (b2);
\coordinate[right=of b2] (bp);
\coordinate[above=of bp] (ap);
\coordinate[left=1.1cm of ap,label=above:$v_{t_1}$] (a1);
\draw[particle] (a1) -- (ap);
\draw[particle] (b2) -- (bp);
\draw[gluon] (ap) --  (bp);
\fill[black] (a1) circle (.05cm);
\fill[black] (b2) circle (.05cm);
\end{tikzpicture}\,+
\begin{tikzpicture}[node distance=0.6cm and 0.8cm]
\coordinate[label=below:$v_{t_2}$] (b2);
\coordinate[right=of b2] (bp);
\coordinate[above=of bp,label=below right:$v$,label=above:$v$] (ap);
\coordinate[left=1.1cm of ap,label=above:$v_{t_1}$] (a1);
\coordinate[right=of ap](cc);
\coordinate[above=0.3cm of cc,label=right:$v_0$](cp);
\coordinate[below=0.1cm of cc,label=right:$v_0$](dp);
\draw[particle] (a1) -- (ap);
\draw[particle] (b2) -- (bp);
\draw[gluon] (ap) --  (bp);
\draw[particle2] (ap) sin  (dp);
\draw[particle] (ap) sin  (cp);
\fill[black] (a1) circle (.05cm);
\fill[black] (b2) circle (.05cm);
\draw (cp) circle (.06cm);
\draw (dp) circle (.06cm);
\end{tikzpicture} +
 \begin{tikzpicture}[node distance=0.6cm and 0.8cm]
\coordinate[label=below:$v_{t_2}$] (b2);
\coordinate[right=of b2] (bp);
\coordinate[above=of bp,label=below right:$w$,label=above:$v$] (ap);
\coordinate[left=1.1cm of ap,label=above:$v_{t_1}$] (a1);
\coordinate[right=of ap](cc);
\coordinate[above=0.3cm of cc,label=right:$v_0$](cp);
\coordinate[below=0.1cm of cc,label=right:$w_0$](dp);
\draw[particle] (a1) -- (ap);
\draw[particle] (b2) -- (bp);
\draw[gluon] (ap) --  (bp);
\draw[particle2] (ap) sin  (dp);
\draw[particle] (ap) sin  (cp);
\fill[black] (a1) circle (.05cm);
\fill[black] (b2) circle (.05cm);
\draw (cp) circle (.06cm);
\draw (dp) circle (.06cm);
\end{tikzpicture} +
\begin{tikzpicture}[node distance=0.6cm and 0.8cm]
\coordinate[label=below:$v_{t_2}$] (b2);
\coordinate[right=of b2,label=below:$v$,label=above right:$v$] (bp);
\coordinate[above=of bp] (ap);
\coordinate[left=1.1cm of ap,label=above:$v_{t_1}$] (a1);
\coordinate[right=of bp](cc);
\coordinate[above=0.1cm of cc,label=right:$v_0$](cp);
\coordinate[below=0.3cm of cc,label=right:$v_0$](dp);
\draw[particle] (a1) -- (ap);
\draw[particle] (b2) -- (bp);
\draw[gluon] (ap) --  (bp);
\draw[particle2] (bp) sin  (dp);
\draw[particle] (bp) sin  (cp);
\fill[black] (a1) circle (.05cm);
\fill[black] (b2) circle (.05cm);
\draw (cp) circle (.06cm);
\draw (dp) circle (.06cm);
\end{tikzpicture} \,+
\begin{tikzpicture}[node distance=0.6cm and 0.8cm]
\coordinate[label=below:$v_{t_2}$] (b2);
\coordinate[right=of b2,label=below:$v$,label=above right:$w$] (bp);
\coordinate[above=of bp] (ap);
\coordinate[left=1.1cm of ap,label=above:$v_{t_1}$] (a1);
\coordinate[right=of bp](cc);
\coordinate[above=0.1cm of cc,label=right:$w_0$](cp);
\coordinate[below=0.3cm of cc,label=right:$v_0$](dp);
\draw[particle] (a1) -- (ap);
\draw[particle] (b2) -- (bp);
\draw[gluon] (ap) --  (bp);
\draw[particle2] (bp) sin  (dp);
\draw[particle] (bp) sin  (cp);
\fill[black] (a1) circle (.05cm);
\fill[black] (b2) circle (.05cm);
\draw (cp) circle (.06cm);
\draw (dp) circle (.06cm);
\end{tikzpicture}\\ 
\nn&+\begin{tikzpicture}[node distance=0.6cm and 0.8cm]
\coordinate[label=below:$v_{t_2}$] (b2);
\coordinate[right=of b2,label=below:$v$,label=above right:$v$] (bp);
\coordinate[above=of bp,label=below right:$v$,label=above:$v$] (ap);
\coordinate[left=1.1cm of ap,label=above:$v_{t_1}$] (a1);
\coordinate[right=of bp](cc);
\coordinate[above=0cm of cc,label=right:$v_0$](cp);
\coordinate[below=0.4cm of cc,label=right:$v_0$](dp);
\coordinate[right=of ap](dd);
\coordinate[above=0.4cm of dd,label=right:$v_0$](ep);
\coordinate[below=0cm of dd,label=right:$v_0$](fp);
\draw[particle] (a1) -- (ap);
\draw[particle] (b2) -- (bp);
\draw[gluon] (ap) --  (bp);
\draw[particle] (bp) sin  (dp);
\draw[particle] (bp) sin  (cp);
\draw[particle] (ap) sin  (ep);
\draw[particle] (ap) sin  (fp);
\fill[black] (a1) circle (.05cm);
\fill[black] (b2) circle (.05cm);
\draw (cp) circle (.06cm);
\draw (dp) circle (.06cm);
\draw (ep) circle (.06cm);
\draw (fp) circle (.06cm);
\end{tikzpicture}\, 
+\begin{tikzpicture}[node distance=0.6cm and 0.8cm]
\coordinate[label=below:$v_{t_2}$] (b2);
\coordinate[right=of b2,label=below:$v$,label=above right:$w$] (bp);
\coordinate[above=of bp,label=below right:$v$,label=above:$v$] (ap);
\coordinate[left=1.1cm of ap,label=above:$v_{t_1}$] (a1);
\coordinate[right=of bp](cc);
\coordinate[above=0cm of cc,label=right:$w_0$](cp);
\coordinate[below=0.4cm of cc,label=right:$v_0$](dp);
\coordinate[right=of ap](dd);
\coordinate[above=0.4cm of dd,label=right:$v_0$](ep);
\coordinate[below=0cm of dd,label=right:$v_0$](fp);
\draw[particle] (a1) -- (ap);
\draw[particle] (b2) -- (bp);
\draw[gluon] (ap) --  (bp);
\draw[particle] (bp) sin  (dp);
\draw[particle] (bp) sin  (cp);
\draw[particle] (ap) sin  (ep);
\draw[particle] (ap) sin  (fp);
\fill[black] (a1) circle (.05cm);
\fill[black] (b2) circle (.05cm);
\draw (cp) circle (.06cm);
\draw (dp) circle (.06cm);
\draw (ep) circle (.06cm);
\draw (fp) circle (.06cm);
\end{tikzpicture}\, +
\begin{tikzpicture}[node distance=0.6cm and 0.8cm]
\coordinate[label=below:$v_{t_2}$] (b2);
\coordinate[right=of b2,label=below:$v$,label=above right:$v$] (bp);
\coordinate[above=of bp,label=below right:$v$,label=above:$w$] (ap);
\coordinate[left=1.1cm of ap,label=above:$v_{t_1}$] (a1);
\coordinate[right=of bp](cc);
\coordinate[above=0cm of cc,label=right:$v_0$](cp);
\coordinate[below=0.4cm of cc,label=right:$v_0$](dp);
\coordinate[right=of ap](dd);
\coordinate[above=0.4cm of dd,label=right:$w_0$](ep);
\coordinate[below=0cm of dd,label=right:$v_0$](fp);
\draw[particle] (a1) -- (ap);
\draw[particle] (b2) -- (bp);
\draw[gluon] (ap) --  (bp);
\draw[particle] (bp) sin  (dp);
\draw[particle] (bp) sin  (cp);
\draw[particle] (ap) sin  (ep);
\draw[particle] (ap) sin  (fp);
\fill[black] (a1) circle (.05cm);
\fill[black] (b2) circle (.05cm);
\draw (cp) circle (.06cm);
\draw (dp) circle (.06cm);
\draw (ep) circle (.06cm);
\draw (fp) circle (.06cm);
\end{tikzpicture}\,+
\begin{tikzpicture}[node distance=0.6cm and 0.8cm]
\coordinate[label=below:$v_{t_2}$] (b2);
\coordinate[right=of b2,label=below:$v$,label=above right:$w$] (bp);
\coordinate[above=of bp,label=below right:$v$,label=above:$w$] (ap);
\coordinate[left=1.1cm of ap,label=above:$v_{t_1}$] (a1);
\coordinate[right=of bp](cc);
\coordinate[above=0cm of cc,label=right:$w_0$](cp);
\coordinate[below=0.4cm of cc,label=right:$v_0$](dp);
\coordinate[right=of ap](dd);
\coordinate[above=0.4cm of dd,label=right:$w_0$](ep);
\coordinate[below=0cm of dd,label=right:$v_0$](fp);
\draw[particle] (a1) -- (ap);
\draw[particle] (b2) -- (bp);
\draw[gluon] (ap) --  (bp);
\draw[particle] (bp) sin  (dp);
\draw[particle] (bp) sin  (cp);
\draw[particle] (ap) sin  (ep);
\draw[particle] (ap) sin  (fp);
\fill[black] (a1) circle (.05cm);
\fill[black] (b2) circle (.05cm);
\draw (cp) circle (.06cm);
\draw (dp) circle (.06cm);
\draw (ep) circle (.06cm);
\draw (fp) circle (.06cm);
\end{tikzpicture},\\
\nn=& \yp_I^2 \,G_{\yp}(t_1,0) G_{\yp}(t_2,0) + \int \dd t' \left[G_{\yp}(t_1,t') G_{\yp}(t_2,t')\left\{\kappa_2+\alpha \yp_I^2 G_{\yp}(t',0)^2 + \alpha \yp_I \zp_I G_{\yp}(t',0)G_{\zp}(t',0) \right\}^2\right],\\
=& \yp_I^2 e^{\diel_2 t_1}e^{ \diel_2 t_2}+ \int \dd t' \left[ e^{\diel_2(t_1-t')}e^{\diel_2(t_2-t')}\left\{\kappa_2 + \alpha \yp_I^2 \left(e^{\diel_2 t'}\right)^2 + \alpha \yp_I \zp_I e^{\diel_2 t'} e^{\diel_3 t'} \right\}^2        \right],
\end{align}
\end{widetext}
where the definitions of the parameters $v_I, \alpha, \kappa_2,...$ are discussed in Eqs.~\eqref{eq-transformQ}-\eqref{eq-transformQ2}.


%

\end{document}